\RequirePackage{fix-cm}
\documentclass[twocolumn]{svjour3}          
\smartqed  

\usepackage{graphicx}
\usepackage{bbding}   

\newcommand*{\affaddr}[1]{#1} 
\newcommand*{\affmark}[1][*]{\textsuperscript{#1}}

\usepackage{algorithm}
\usepackage{algpseudocode}

\usepackage{amsmath, amssymb}
\usepackage{graphicx}
\usepackage{bm}

\usepackage{cuted}
\usepackage{algorithmicx}

\usepackage{multirow}

\usepackage{booktabs}

\usepackage{xparse}

\usepackage{pgffor}
\usepackage{etoolbox}

\usepackage{float}

\usepackage{enumitem}

\usepackage{colortbl}
\usepackage{caption}
\usepackage{subcaption}
\usepackage{xspace}
\usepackage{hyperref}

\usepackage{arydshln}

\newcommand{\revision}[1]{{\color{black}#1}}

\newcommand{\ie}{\textit{i}.\textit{e}.} 

\usepackage{amsthm}
\newtheorem{thm}{Statement}

\let\oldref=\ref
\renewcommand{\ref}[1]{\textcolor{green}{\oldref{#1}}}

\let\oldcite=\cite
\renewcommand{\cite}[1]{\textcolor{cyan}{\oldcite{#1}}}

\begin{document}

\title{PoseAlign: Sculpting Pose-Consistent Meshes via Text-Guided Deformation
}


\author{Shijin Wang   \protect\affmark[1]      \and
        Zichong Chen  \protect\affmark[1]     \and
        Yang Zhou     \protect\affmark[1]    \and
        Hui Huang     \protect\affmark[1]     
}

\authorrunning{S. Wang, Z. Chen, Y. Zhou, and H. Huang} 

\institute{
        \Envelope \hspace{0.01cm} Yang Zhou  \\
        \hspace*{0.26cm} zhouyangvcc@gmail.com \\
        \\
        \affaddr{\affmark[1] 
        Guangdong Provincial Key Lab. of Visual Media and Multidim. Intelligence, CSSE, Shenzhen University, China.} 
}


\maketitle

\begin{abstract}

Mesh deformation, the process of altering the vertex positions of a 3D mesh while preserving its topological structure, is a cornerstone of computer graphics. Despite the recent emergence of numerous text-guided 3D mesh deformation methods, deforming an initial mesh into one that both adheres to text prompts and preserves its pose remains challenging. This paper proposes PoseAlign, which decomposes text-guided mesh deformation into two stages: global pose scaling and local detail sculpting. Specifically, in the first stage, we introduce the Laplacian as a differentiable mesh representation to enable more efficient yet smoother global deformation. Then, we propose a novel pose-aligned SDS loss by adapting score distillation sampling (SDS) with an attention-sharing mechanism, which sculptures fine-grained geometric details for the deformed mesh while preserving its original pose. PoseAlign significantly enhances the controllability of the overall deformation process, achieving a favorable balance between pose preservation and text alignment. Experiments demonstrate the competitive advantages of our method in text alignment and mesh quality. Code is available at: \textcolor{magenta}{\url{https://cousingrade6.github.io/PoseAlign/}}.

\keywords{Mesh deformation \and Geometry representation \and Pose-aligned SDS Loss \and Attention Mechanism}
\end{abstract}


\section{Introduction}

\begin{figure}[!t]
     \includegraphics[width=1.0\linewidth]{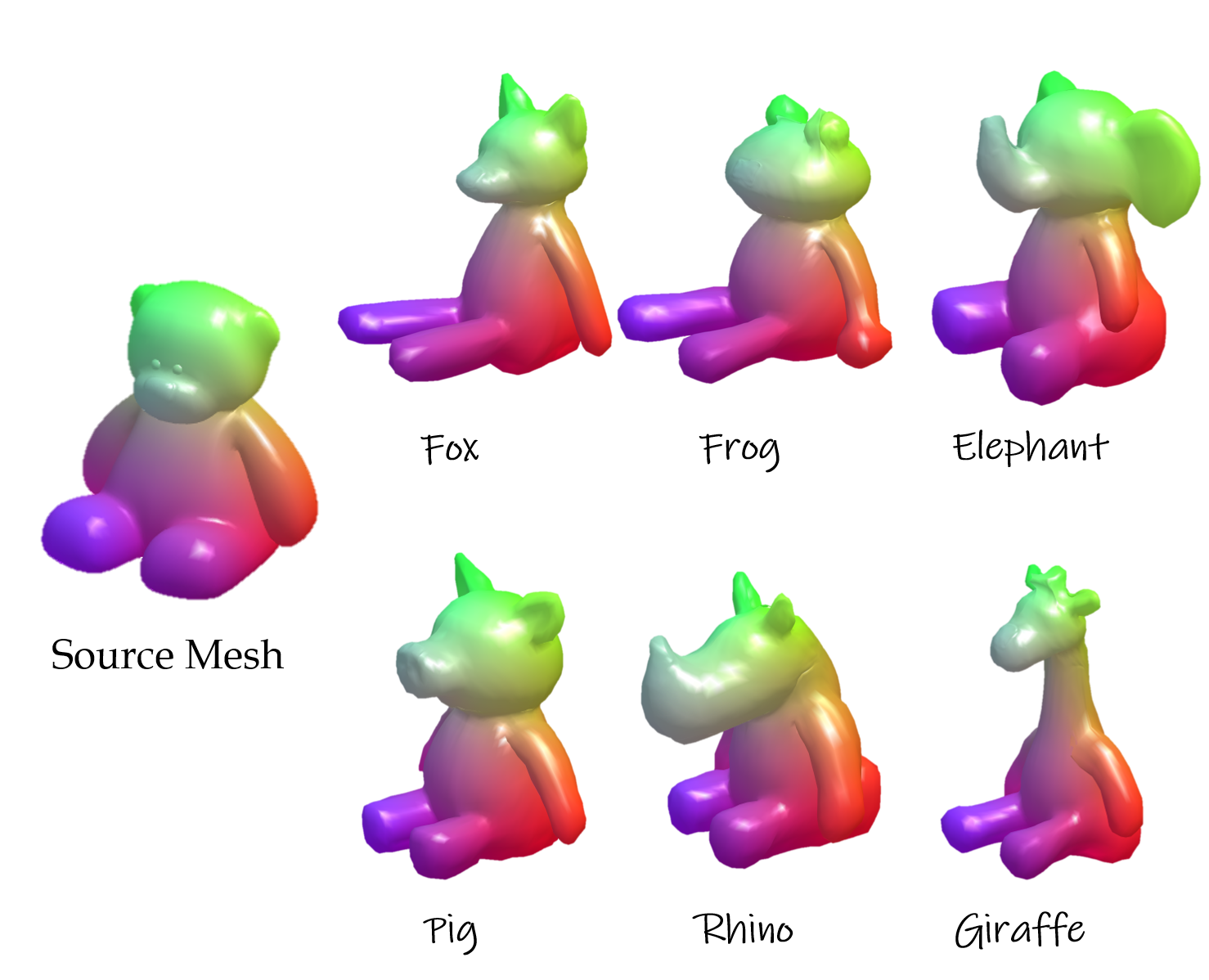}
     \centering
      \caption{\textbf{Pose-preserved text-guided mesh deformation.} Given a source mesh, PoseAlign is capable of deforming it to a target mesh that aligns with the text semantics while preserving its original pose.}
    \label{fig:teaser}
\end{figure}

Mesh deformation is a fundamental task in computer graphics with broad applications in content creation, character posing, morphing, and animation \cite{intro:corman2019functional-character,intro:dodik2023vbc,intro:gal2009iwires,intro:mitra2007dgr}. Traditional methods rely on geometric priors (e.g, As-Rigid-As-Possible \cite{arap}, Laplacian deformation \cite{laplacian}), or handle-based controls \cite{intro:jacobson2013Algorithms,intro:Sorkine2009interactive}, which require significant manual efforts to achieve plausible results. 

The advent of Vision-Language Models (VLMs) \cite{vlm:lu2019bert,radford2021clip,vlm:li2022blip} and diffusion models \cite{Rombach2021HighResolutionIS,Podell2023SDXLIL,deepfloyd} has revolutionized this domain by enabling text-guided deformation, where natural language prompts intuitively drive geometric manipulation. Moreover, per-face Jacobian \cite{aigerman2022neural} emerged as a pivotal differential representation, replacing direct vertex displacements to prioritize smoother deformations. Text-driven methods leverage differentiable rendering and pretrained models (e.g., the CLIP~\cite{radford2021clip} or diffusion models~\cite{Rombach2021HighResolutionIS}) to connect geometry with textual semantics, transforming deformation into an optimization problem guided by text-image alignment.

\begin{figure}[t]
    \centering
    \includegraphics[width=1.0\linewidth]{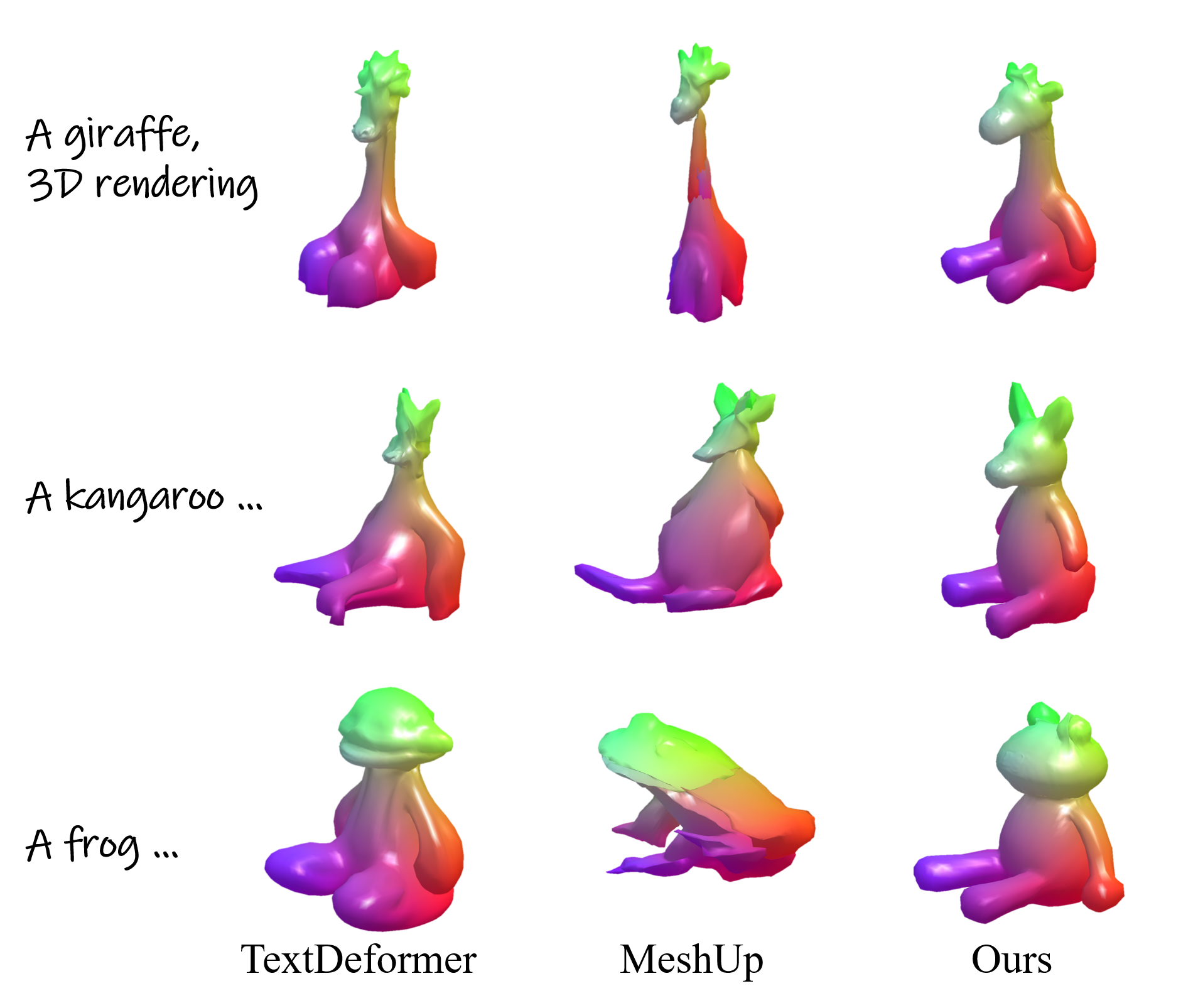}
    \caption{\textbf{Motivation of PoseAlign.} Given the source mesh displayed in Fig. \ref{fig:teaser}, we aim to deform it to conform to user-specified textual descriptions. Previous techniques, such as TextDeformer and MeshUp, exhibit severe limitations in generating high-fidelity meshes that simultaneously preserve the source mesh's pose and achieve semantic alignment with the input text. Our approach builds upon these observations and yields meshes with significantly improved quality, more precise pose preservation, and enhanced text fidelity.}
    \label{fig:motivation}
\end{figure}

However, current methods \cite{gao2023text-deformer,xu2024fusion-deformer,kim2024meshup} often fall short in two aspects: large-scale deformation quality and pose preservation. As shown in Fig. \ref{fig:motivation}, TextDeformer \cite{gao2023text-deformer} partially preserves the pose of the input source mesh but struggles to align with the text description and yields a suboptimal deformation quality. In contrast, MeshUp \cite{kim2024meshup} successfully deforms the source mesh to align with the target object, yet completely loses the original pose and introduces noticeable artifacts.

We attribute them to Jacobian Independence and Visual Prior Bias. On the one hand, the Jacobians are computed independently, meaning that adjacent faces are agnostic to changes in one another during optimization, resulting in incoherent deformation between them. A Jacobian regularization is thus needed to suppress over-deformation. On the other hand, vision-language and diffusion models prioritize common poses (e.g., qua-drupedal standing animals) due to training data bias. When deforming a sitting teddy bear, the mesh may override the pose to match learned priors, even if this introduces catastrophic distortion.

\begin{figure}[t]
    \centering
    \includegraphics[width=1.0\linewidth]{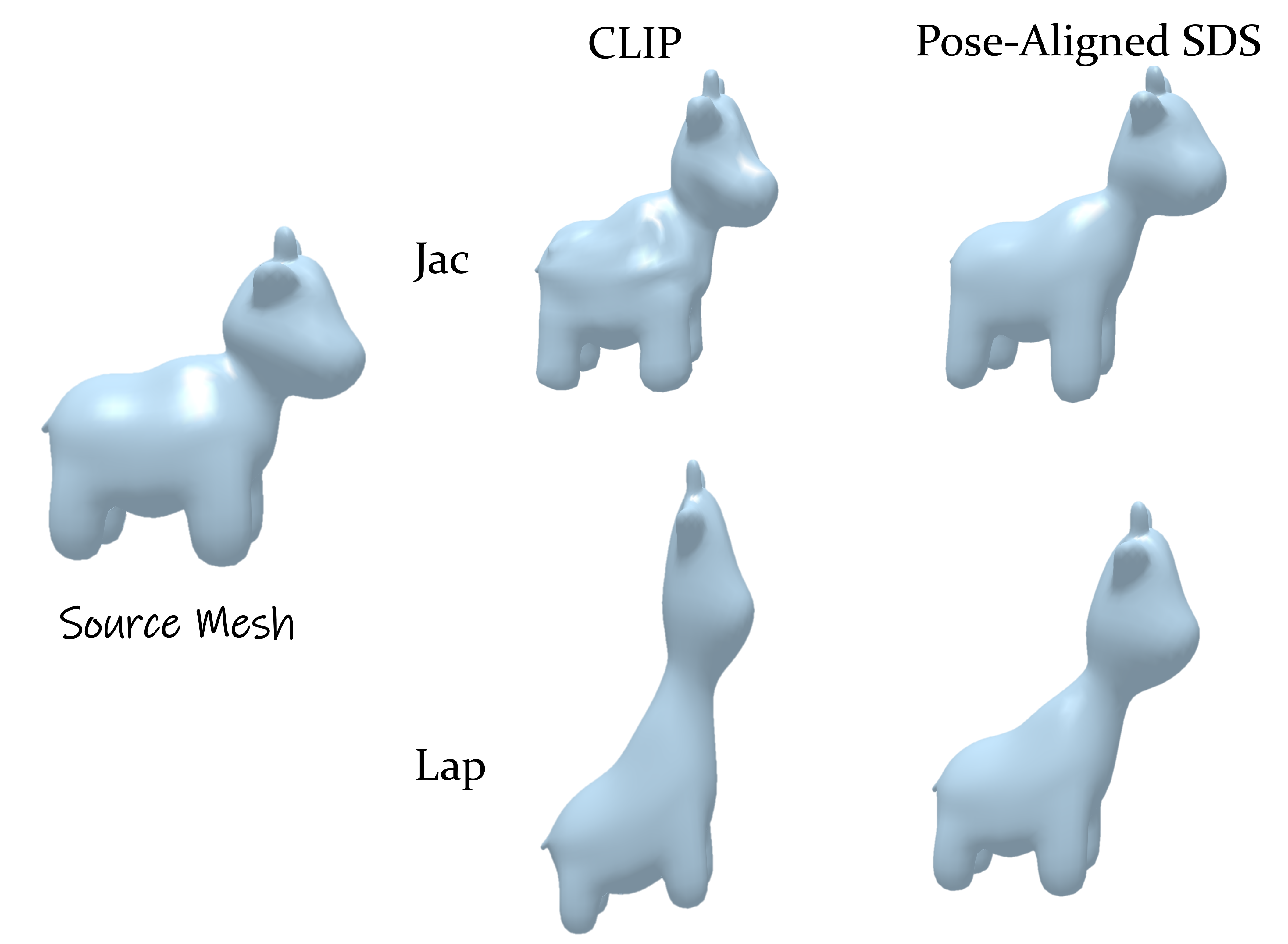}
    \caption{\textbf{Jacobians vs Laplacians and CLIP Loss vs SDS Loss.} Meshes deformed using different geometric representations and loss functions from the same source mesh into ``\texttt{a giraffe}'' with 200 iterations.  In terms of loss functions, CLIP loss is more computationally efficient, while it performs significantly worse in terms of smoothness than SDS loss with attention sharing. In terms of geometric representations, Laplacian-based optimization produces more pronounced stretching effects and better suppresses artifacts.}
    \label{fig:jac-vs-lap}
\end{figure}

To address these challenges, we propose PoseAlign, a novel method that decomposes the text-driven mesh deformation process into two distinct stages: global pose scaling and local detail sculpting. The global pose scaling stage deforms the input mesh's contour to approximately conform to the input prompt. Specifically, we introduce the Laplacian as a differential parameter and optimize it with CLIP loss. Using CLIP loss is motivated by its inherent encoding of global structural priors and high computational efficiency, which make it well-suited to large-scale deformations. However, noisy gradients of the CLIP loss can induce high-frequency artifacts, as shown in the results of TextDeformer in Fig.~\ref{fig:motivation}. To mitigate this, we leverage the Laplacian to preserve local smoothness throughout the deformation process. Then, at the local detail sculpting stage, we refine the mesh's fine-grained features, thereby enhancing its alignment with the input prompt at a finer scale. During this phase, per-face Jacobian is accordingly updated. Critically, preserving the pose of the input mesh remains a key requirement. Our core insight is that 
diffusion models have a strong semantic generalization ability, and their self-attention features can identify the same semantic component across objects and perspectives.
Building on this capability, we adapt SDS loss~\cite{poole2022sds} by incorporating an attention sharing mechanism~\cite{hertz2024stylealigned}. 
Semantic correspondences between the input and output meshes are maintained during the whole deformation process, and therefore, the original pose is preserved. Qualitative and quantitative results show that, compared with existing methods, PoseAlign overcomes the common issues in large-scale deformation quality and pose preservation, offering a more robust solution for text-driven mesh deformation.

Our contributions are summarized as follows:
\begin{itemize}
    \item We introduce PoseAlign, a new text-guided mesh deformation method that enables a more controllable deformation process and outputs meshes that are both text-aligned and pose-preserved.
    \item We explore the potential of Laplacian as a differentiable mesh representation, achieving more efficient and smoother deformation. 
    \item We propose a novel pose-aligned SDS loss that leverages the attention-sharing mechanism of diffusion models, thereby retaining the input mesh's pose in the output.
    \item We demonstrate the superiority of our approach through qualitative and quantitative comparisons with existing methods and compatibility with downstream applications.
\end{itemize}


\section{Related Work}


\subsection{Diffusion-Based Image Editing}
Recent advancements in text-to-image (T2I) diffusion models, particularly within the Stable Diffusion series \cite{esser2024sd3,Podell2023SDXLIL,deepfloyd}, have revolutionized image editing, significantly enhancing both the quality and diversity of images generated from text prompts. Currently, the predominant diffusion-based image editing methods can be broadly categorized by their research focus in image-to-image translation \cite{relate:i2i:meng2022sdedit,relate:i2i:parmar2023zeroshotimagetoimagetranslation,relate:i2i:ye2023ip-adapter}, local image editing \cite{relate:local:couairon2022diffedit,relate:local:huberman2024edit,relate:local:shi2023dragdiffusion}, style transfer \cite{relate:st:he2024freestyle,relate:st:wang2023stylediffusion,relate:st:zhang2023inst}, personalization \cite{relate:person:alaluf2023neti,relate:person:textualinversion,relate:person:voynov2024p}, and other related directions. From a training methodology perspective, these approaches are further classified into pretrained (introducing new pretrained modules \cite{relate:pretrained:gao2024styleshot,relate:pretrained:wang2024styleadapter,relate:pretrained:xing2024csgo}, finetuning (by finetuning diffusion model \cite{relate:ft:Ruiz2022DreamBoothFT,relate:ft:chen2025styleblend,relate:ft:jones2024customizingtexttoimagemodelssingle}), and training-free methods \cite{relate:tf:AttDistill25,relate:tf:Chung_2024_CVPR,relate:tf:rout2024rbmodulation,relate:tf:xu2025stylessp}. Training-free methods, which inject reference features into the denoising process or manipulate attention features, can fully leverage the prior knowledge learned in pretrained diffusion models to perform image editing without requiring additional training. Owing to their inherent ease of use, these training-free methods can be rapidly adapted and applied to the 3D manipulation task.

\subsection{3D Generation} 
Due to the remarkable success of diffusion and flow-based models in 2D image generation, text-to-3D methods have attracted increasing attention from researchers. 3D data can be represented through diverse modalities, encompassing explicit formats (e.g., meshes, voxel grids, point clouds) and implicit representations (e.g., NeRFs \cite{mildenhall2020nerf}, Gaussian Splatting \cite{kerbl3Dgaussians}). Among these, meshes are the de facto standard in 3D graphics pipelines for their native support for intuitive geometric operations and rendering. Some prior work has attempted to encode the shape information of meshes into a codebook and decode new meshes using autoregressive mechanisms \cite{relate:md:meshgpt_2024,yin2023shapegpt,relate:md:alliegro2023polydiff,chen2024meshanything,chen2024meshanythingv2artistcreatedmesh}. 
In addition, many methods introduce a VAE architecture \cite{vae} to encode 3D objects into compact latent codes for the diffusion \cite{3DShape2VecSet,zhao2023michelangelo,li2024craftsman,zhang2024clay,relate:t23d:yang2024hunyuan3d} or flow matching \cite{relate:t23d:hunyuan3d22025tencent,relate:t23d:xiang2024trellis} process, and decode them to signed distance field, thereby producing meshes. Building upon these advancements, several approaches have recently been proposed leveraging state-of-the-art 3D generative models for mesh editing \cite{relate:md:wukong,relate:md:lee2025revived,relate:md:zhou2025anchorflow}. 
While these text-to-3D generation-based methods have achieved remarkable progress, they primarily address the problem of creating novel 3D assets from scratch. However, in numerous professional graphics pipelines and content-creation scenarios, the core requirement is to edit and stylize existing 3D assets while preserving their topology and structural properties. This distinction motivates our focus on text-guided mesh deformation rather than unconditional generation. 

Therefore, we argue that text-guided mesh deformation solves a distinct and valuable problem: the controlled, semantically aware stylization of existing 3D models without sacrificing their structural integrity and predefined attributes. This work, PoseAlign, is designed specifically to address the core challenge within this paradigm: balancing strong semantic alignment with the text prompt against faithful preservation of the source mesh’s pose.

\subsection{Mesh Deformation}
Early mesh deformation methods primarily relied on user-controlled handles to define a deformation space.
Energy-based methods, such as ARAP \cite{arap} and Laplacian surface editing \cite{laplacian}, minimize deformation energies to preserve geometric details while allowing users to manipulate control points. 
Skinning techniques interpolate vertex coordinates using weights associated with handles \cite{relate:md:Fulton2018,relate:md:skinning2014}, enabling the natural articulation of deformable objects such as human figures. Complementary methods include cage-based \cite{relate:md:Yifan2020NeuralCage,relate:md:Michel2023cage,relate:md:Lin2024cage} and skeleton-based approaches \cite{relate:md:Automaticrigging,relate:md:Weber2007ContextAwareSS}. 
However, all these traditional methods are constrained: they depend on manual handles and lack semantic understanding.  

Recent efforts leverage VLMs to encode text and directly drive deformations, thus bypassing extensive manual input. For example, some methods use diffusion-based or CLIP-based objectives to guide Jacobian-based representations and, in turn, generate deformation results that align with user input text \cite{gao2023text-deformer,xu2024fusion-deformer,relate:md:wang2024headevolver}. By introducing the rich image priors of diffusion models, mesh deformation tasks can now be integrated with various diffusion-based image editing frameworks, thereby enabling advanced applications such as realistic handle-based deformations \cite{relate:md:yoo2024apap,relate:md:xie2023dragd3d}, concept combination \cite{kim2024meshup}, and geometric stylization \cite{dinh2025geometryinstyle}. 
In this work, our main goal is to extend these techniques to achieve both text-aligned and pose-preserved deformation processes and high-fidelity outputs.

\section{Preliminaries}

\subsection{Jacobian-Based Mesh Deformation}
Given an input mesh $\mathcal{M}=\{\mathcal{V}, \mathcal{F}\}$, the Jacobian matrix $J_i \in \mathbb{R}^{3\times3}$ represents the local linear transformation between the undeformed and deformed states of the $i$th triangle, which is defined as the gradient of the deformation map $\mathbf{\phi}$ evaluated on face $f_i$, \ie, $J_i=\nabla_i\mathbf{\phi}$, where $\nabla_i$ denotes the spatial gradient operator on $f_i$.
The Jacobian $J_i$ captures the affine transformations (including rotation, scaling, shear) applied to $f_i$. To compute the deformed vertex positions $\mathcal{V^*}$ from the Jacobians, a Poisson equation introduced by \cite{aigerman2022neural} is solved:
\begin{equation}
    \phi^* = \arg\min_\mathbf{\phi} \sum_{f_i\in\mathcal{F}} |f_i| \|\nabla_i(\mathbf{\phi})-J_i\|^2_2=\Delta^{-1}\mathcal{A}\nabla^{\top}\mathcal{J},
    \label{eq:poisson}
\end{equation}
where $|f_i|$ is the area of $f_i$, $\Delta$ is the cotangent Laplacian operator, $\mathcal{A}$ is the face mass matrix, and $\mathcal{J}$ is the concatenation of all Jacobians. This global optimization enforces compatibility among faces and ensures that the vertex positions remain faithful to the target Jacobian $\mathcal{J}$ while preserving mesh continuity.

\subsection{Text-Guided Mesh Deformation via SDS}
The foundational principle of text-guided mesh deformation using diffusion models is to leverage pretrained text-to-image (T2I) diffusion models to access the plausibility of 3D mesh renderings conditioned on a text prompt. Formally, given a text prompt $y$ and a pretrained T2I diffusion model with a denoising U-Net $\epsilon_\eta$, the objective is to optimize a 3D mesh parameterized by $\theta$, such that its rendered image $\mathbf x=g(\theta)$ aligns with the image distribution induced by the T2I diffusion model. This optimization is achieved by first sampling a noise $\epsilon$ and perturbing the rendered image to obtain $\mathbf{x}_t$, which is then fed into the denoising U-Net $\epsilon_\eta$ to predict the target noise. A diffusion loss between the sampled and predicted noise is used to compute gradients, which are then backpropagated to update the mesh parameters $\theta$. To reduce computational overhead and guided the optimization more effectively, Score Distillation Sampling (SDS) proposed by \cite{poole2022sds} formalizes this process as 
\begin{equation}
    \nabla_\theta\mathcal{L}_\text{SDS}=\mathbb{E}_{t,\epsilon\sim\mathcal{N}(0, 1)}\Bigg[\omega(t)(\epsilon_\eta(\mathbf{x}_t;y,t)-\epsilon)\frac{\partial \mathbf{x}}{\partial\theta}\Bigg],
    \label{eq:sds}
\end{equation}
where $\omega(t)$ is a weighting function with $t \sim U(0.02, 0.98)$, and $\mathbf x_t$ denotes the rendered image perturbed with noise at timestep $t$. 

\subsection{Self-Attention in T2I Diffusion Models}
Our approach builds upon DeepFloyd-IF XL \cite{deepfloyd}, a T2I diffusion model that operates directly in RGB space and uses a U-Net architecture with multiple convolutional blocks and transformer blocks \cite{vaswani2017attention} as its denoising network. Prior studies have shown that both the convolutional and attention features within the U-Net can serve as highly informative representations for capturing image semantic and structural cues \cite{pnpDiffusion2022,cao2023masactrl,hertz2024stylealigned,alaluf2024cross}. In this work, we focus specifically on the self-attention features and introduce an attention-sharing mechanism for pose alignment during mesh deformation, inspired by \cite{hertz2024stylealigned}. The self-attention operation is defined as
\begin{equation}
\text{Self-Attention}(Q,K,V)=\text{softmax}{\big(}\frac{QK^T}{\sqrt{d}}{\big)}\cdot V,
\label{eq:attention}
\end{equation}
where $d$ denotes the feature dimension, and $Q$, $K$, and $V$ are the Query, Key, and Value, respectively. 


\section{Method}

\begin{figure*}[!t]
  \centering
  \includegraphics[width=0.8\linewidth]{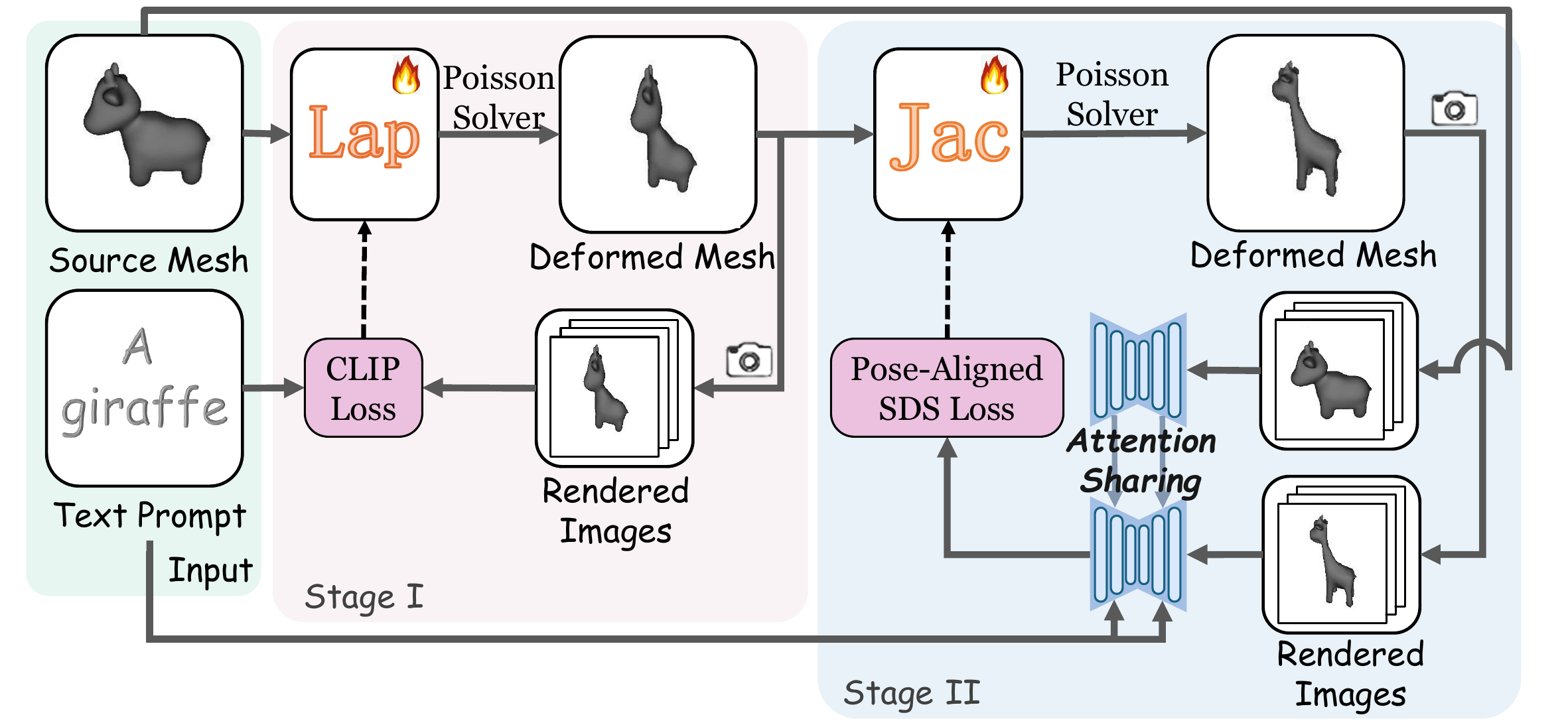} 
  \caption{\textbf{Method Pipeline.} We decompose the mesh deformation process into two stages:  Global Pose Scaling and Local Detail Sculpting. The first stage optimizes the Laplacians to generate a deformed mesh via Poisson solver, followed by differentiable rendering \cite{Laine2020nvdiffrast} to produce rendered images. These images are then evaluated against the text prompt using the CLIP loss. The second stage leverages the Jacobians to further optimize the fine details of the deformed mesh from Stage I. Using the Poisson solver and differentiable rendering again, the updated mesh is supervised by the Pose-Align SDS loss, with the attention sharing mechanism incorporated to ensure coherent geometric and textural details.  }
  \label{fig:pipeline}
\end{figure*}

In this section, we introduce PoseAlign, a two-stage mesh deformation approach consisting of Global Pose Scaling (Sec. \ref{sec:global}) and Local Detail Sculpting (Sec. \ref{sec:local}). Fig.~\ref{fig:pipeline} illustrates the pipeline of our proposed method. Given an input mesh, our method deforms the mesh to align with user-provided text prompts while preserving its original pose. 
 
\subsection{Global Pose Scaling}
\label{sec:global}
In the first stage, our objective is to perform efficient global geometric scaling of the input mesh while preserving its details and original pose. CLIP is well-suited for large-scale deformation due to its efficiency and capacity for global semantic alignment. 
Nevertheless, it is incompatible with the Jacobians, as it lacks detailed geometric priors, whereas Jacobians specialize in manipulating local geometry. Using CLIP directly with Jacobians introduces many artifacts, as shown in Fig. \ref{fig:jac-vs-lap}. 
To address this limitation, we introduce mesh Laplacian coordinates, a differential-geometric representation that encodes intrinsic surface structure. Lap-lacian based optimization enhances stretching and reduces artifacts introduced by CLIP loss, as illustrated in Fig \ref{fig:jac-vs-lap}. By optimizing the Laplacian with the CLIP loss, we strike a balanced trade-off between highly efficient semantic alignment and global geometric smoothness.

Generally speaking, the Laplacian of a mesh vertex $v_i$, denoted $\delta_i$, is a fundamental quantity in geometry processing that encodes the local geometric context of $v_i$ relative to its neighborhood. Formally, for a triangular mesh $\mathcal{M}=\{\mathcal{V}, \mathcal{F}\}$, the discrete Laplacian coordinate $\delta_i\in\mathbb R^{3}$ of $v_i$ is defined as:
\begin{equation}
    \delta_i=v_i-\frac{1}{N(i)}\sum_{j\in\mathcal{N}(i)}w_{ij}v_j,
    \label{eq:laplacian}
\end{equation}
where $\mathcal{N}(i)$ denotes the set of neighboring vertices of $v_i$, $N(i)=|\mathcal{N}(i)|$ is the degree of $v_i$, and $w_{ij}$ are cotangent weights \cite{Pinkall1993cotan}
that enforce geometric smoothness. 
By optimizing the Laplacian coordinates $\mathcal{L}$ instead of the per-face Jacobian $\mathcal{J}$ or the vertex positions $\mathcal{V}$ directly, we inherently impose a local smoothness prior. The gradient $\nabla \mathcal{L}_{\text{CLIP}}$ updates the deviation vector $\delta_i$, which represents the offset of a vertex from its neighborhood centroid. The noisy update introduced by the CLIP loss, which attempts to move a vertex disproportionately toward its neighbors, would result in large, unstable changes in $\delta_i$, which the optimization naturally discourages, thereby promoting coherent local motion.

By introducing the Laplacian, the Poisson equation can be rewritten as 
\begin{equation}
    \phi^*=\arg\min_\phi\sum_{i}\|\Delta_i(\phi)-L_i \|^2_2=\Delta^{-1}\mathcal{L},
    \label{eq:lap_poisson}
\end{equation}
where $\Delta_i$ is the Laplacian operator on vertex $v_i$ and $\mathcal{L}$ is a concatenation of $L_i$. In this case, $\phi^*$ functions as the deformation map and its Laplacian $\Delta_i(\phi)$ approximates the target Laplacian $\mathcal{L}$, which is optimized by maximizing the cosine similarity between the CLIP embeddings of text prompts and rendered images.

\subsection{Local Detail Sculpting}\label{sec:local}

\begin{figure*}[!t]
    \centering
    \includegraphics[width=0.88\linewidth]{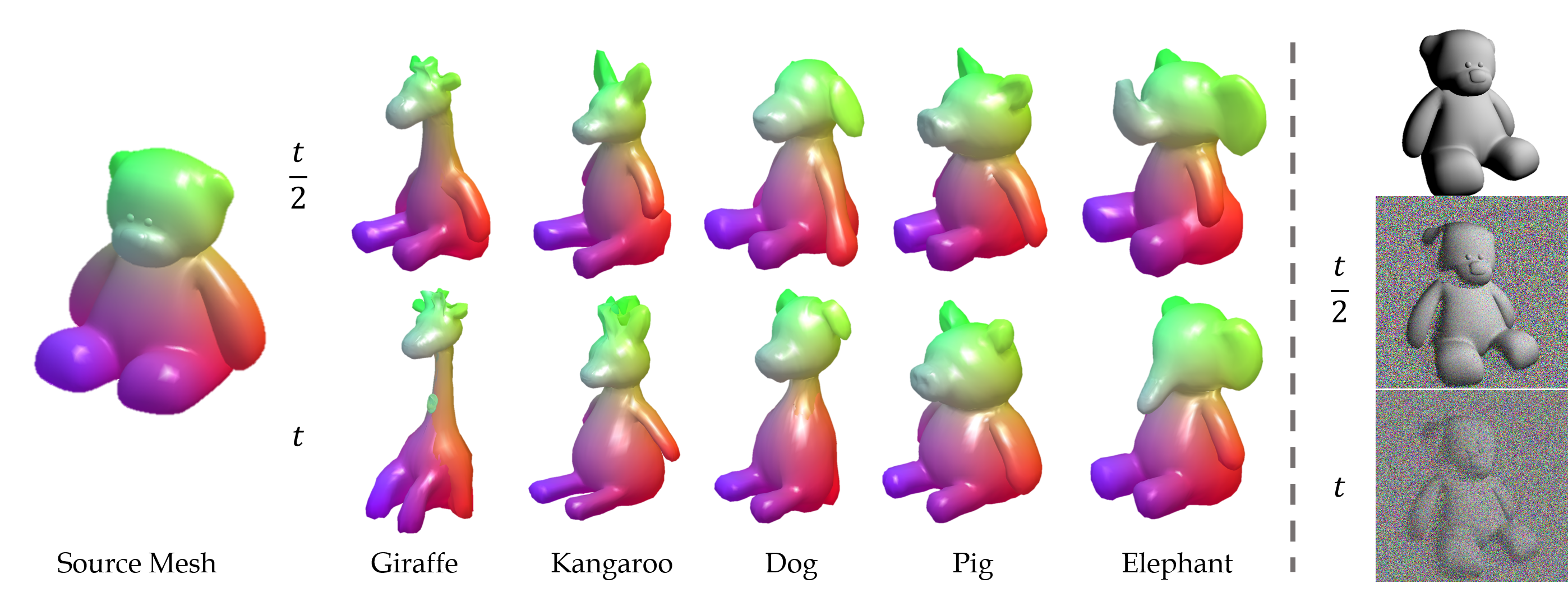}
    \caption{\textbf{Comparison of the noise timestep.} The left comparison shows the deformed results when using timestep $t/2$ (top row) and $t$ (bottom row) for the reference branch. The right column presents the corresponding noisy rendered images of the source mesh at these timesteps, illustrating the difference in feature preservation.}
    \label{fig:ablation_t}
\end{figure*}

In the second stage, our goal is to refine the fine-grained geometric features of the deformed mesh while ensuring it remains aligned with the original pose. To this end, we adopt Jacobians as the geometric representation and the SDS loss as the optimization objective. The Jacobians enable fine-grained manipulation of mesh faces and promote locally plausible geometric structures, while the SDS loss provides more faithful semantic guidance grounded in the diffusion model distribution, as demonstrated in Fig.~\ref{fig:jac-vs-lap}. However, a challenge remains at this stage: maintaining alignment with the original mesh pose during optimization. To address this issue, we introduce an attention-sharing mechanism inspired by \cite{hertz2024stylealigned} into the score distillation, which facilitates pose alignment and further stabilizes the optimization process. 

\begin{figure*}[!t]
    \centering
    \includegraphics[width=1.0\linewidth]{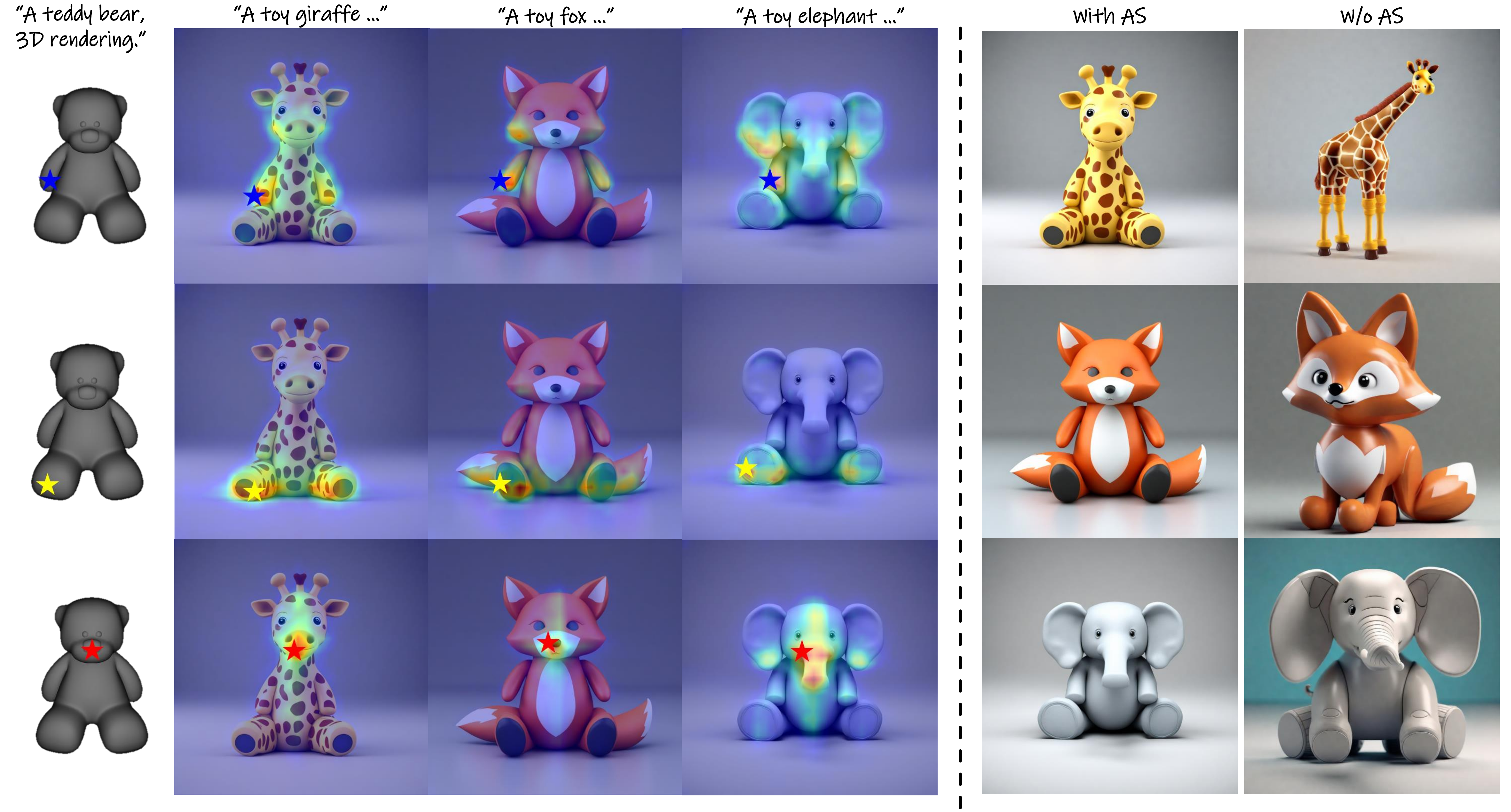}
    \caption{\textbf{Semantic correspondence captured by attention-sharing mechanism.} Left: We visualize the self-attention probability maps across the generated images, with distinct colored stars marking different semantic components. By introducing the attention-sharing (AS) mechanism, the diffusion model effectively captures semantic correspondences from the rendered reference image (leftmost) across the generated images, thereby enabling pose alignment across different generation processes. Right: Images generated by the primordial Diffusion model (rightmost) fail to preserve the input pose.}
    \label{fig:attention_map}
\end{figure*}

Specifically, given the deformed mesh $\mathcal{M}$ obtained in Stage I and the source mesh $\mathcal{M}_0$, we first randomly sample a viewpoint and render images $\mathbf{x}$ and $\mathbf{x}_0$ of the two meshes from this viewpoint, respectively. We characterize the pose-aligned denoising process using an adjusted score function $\epsilon(\mathbf{x}_t; y,t,\mathbf{x}_{0,t/2})$, where $\mathbf{x}_t$ and $\mathbf{x}_{0,t/2}$ form a two-branch noisy image input. The reference branch of $\mathbf{x}_{0,t/2}$ is used to extract reference features, which serve as the basis for the attention-sharing mechanism. Here, $\mathbf{x}_t$ and $\mathbf{x}_{0,t/2}$ are obtained by perturbing $\mathbf{x}$ and $\mathbf{x}_0$ with the same noise $\epsilon\sim\mathcal{N}(0,1)$ at timesteps $t$ and $t/2$, respectively. 

\revision{\textbf{Choice of reference timestep.} The use of $t/2$ for the reference branch is motivated by the properties of the diffusion noise schedule. In the forward diffusion process, the noise level $\bar{\alpha}_t$ increases monotonically with $t$. At the full timestep $t$, the reference image $\mathbf{x}_0$ would be corrupted by the same high noise level as the target branch, causing its structural and semantic features to be largely obliterated, which weakens the pose-alignment signal conveyed through attention sharing. We validate this in Fig.~\ref{fig:ablation_t}, where $t/2$ consistently yields better pose preservation than $t$ across multiple test cases. On the other hand, we empirically find that using a very small timestep (e.g., $t/4$) preserves too much low-level detail of $\mathbf{x}_0$, which can overly constrain the target generation and interfere with the semantic deformation driven by the text prompt. The midpoint $t/2$ therefore provides a principled trade-off: it retains sufficient high-level structural information for effective attention-based correspondence, while remaining noisy enough to avoid suppressing the semantic transformation of the target branch.}

In self-attention layers of the denoising network $\epsilon_\eta$, we apply an attention-sharing mechanism, similar to~\cite{hertz2024stylealigned}, for two purposes: i) to align the pose of $\mathcal{M}$ with $\mathcal{M}_0$ and ii) to stabilize the optimization process. More specifically, we first apply AdaIN~\cite{AdaIN} to the $Q$ and $K$ features of $\mathbf{x}$ in the self-attention layers to match them to those of the reference branch ($Q_0$ and $K_0$ of $\mathbf{x}_{0}$), formulated as: 
\begin{equation}
    \hat Q = \text{AdaIN}(Q, Q_0), \hat K = \text{AdaIN}(K, K_0).
\end{equation}
Subsequently, we share the $K_0$ and $V_0$ features from the reference branch to the target branch, and the self-attention operation of the target branch is then formulated as:
\begin{equation}
    \text{Self-Attention}(\hat Q, \text{concat}(K_0, \hat K), \text{concat}(V_0, V)),
    \label{eq:attention-share}
\end{equation}
where $V, V_0$ represents the Value of $\mathbf{x}, \mathbf{x}_{0}$ in self-attention, respectively, and  $\text{concat}(\cdot,\cdot)$ denotes token-wise concatenation. Afterward, we denote the score function with attention-sharing as $\epsilon^{\text{align}}_\eta(\mathbf{x}_t; y,t,\mathbf{x}_{0,t/2})$, which is then used to compute the pose-aligned SDS loss and optimize the Jacobians $\mathcal{J}$ of mesh $\mathcal{M}$ from Stage I via backpropagation. We reformulate the gradient of our pose-aligned SDS loss $\mathcal{L}_\text{SDS}^{\text{align}}$ as follows: 
\begin{equation}
    \begin{aligned}
        &\nabla_\mathcal{J}\mathcal{L}_\text{SDS}^{\text{align}} \\
        &=\mathbb{E}_{t,\epsilon\sim\mathcal{N}(0, 1)}\Bigg[\omega(t)(\epsilon^{\text{align}}_\eta(\mathbf{x}_t; y,t,\mathbf{x}_{0,t/2})-\epsilon)\frac{\partial \mathbf{x}}{\partial\mathcal{J}}\Bigg].
        \label{eq:sds-with-attention}
    \end{aligned}
\end{equation}

\begin{figure*}[t]
    \centering
    \includegraphics[width=1.0\linewidth]{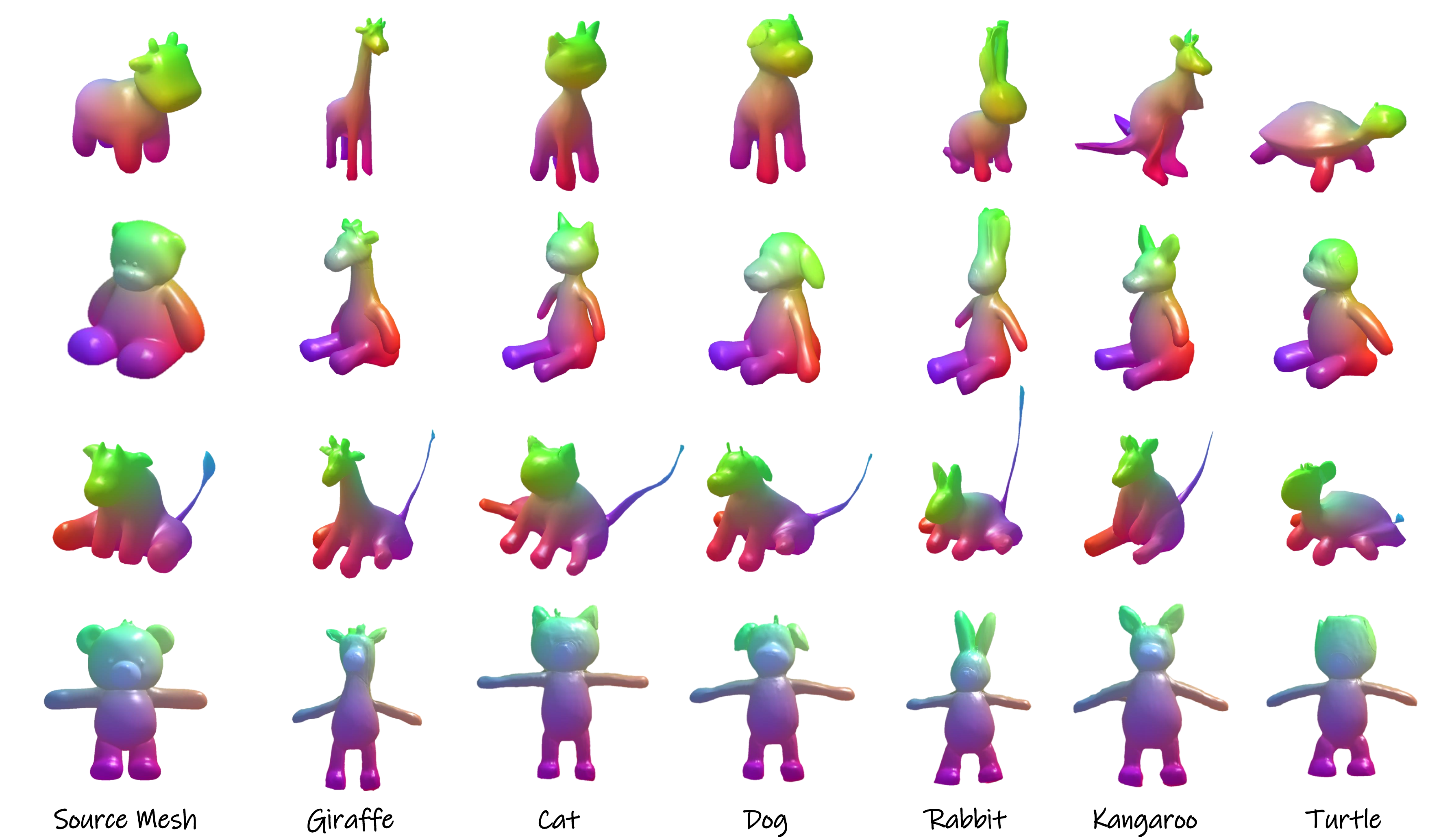}
    \caption{\textbf{Results Gallery of PoseAlign.} PoseAlign robustly preserves the original pose of the source meshes across various animal models, maintaining structural alignment while enabling versatile shape transformations. }
    \label{fig:gallery}
\end{figure*}

To better illustrate the role of the attention-sharing (AS) mechanism within our pose-aligned score distillation, we compare images generated with and without AS. As shown in Fig.~\ref{fig:attention_map}, right. Results obtained with attention-sharing demonstrate clear pose alignment with the reference image, whereas those without AS fail to maintain the original pose. Beyond pose consistency, results with AS also exhibit more coherent and stable backgrounds with reduced appearance inconsistencies. This effectively mitigates visual disturbances during optimization, allowing the model to focus more on geometric deformation. In addition, we visualize the attention maps at several anchor points across the generated images. As shown in Fig.~\ref{fig:attention_map} left, the attention probabilities projected onto the reference image exhibit strong semantic alignment with the spatial locations of the query points, highlighting the model's ability to capture and propagate contextual relationships across distant regions. As a result, both pose alignment and optimization stabilization processes are ensured. 

Critically, the attention-sharing mechanism does more than align image features. It fundamentally constrains the geometric deformation of the 3D mesh. The mechanism establishes dense semantic correspondences between the rendered views of the source mesh $\mathbf{x}_{0}$ and the target mesh $\mathbf{x}$ within the U-Net's latent space. When the denoising model predicts noise based on the mixed attention in Eq.~\ref{eq:attention-share}, it is compelled to generate an image where semantic parts (e.g., head, arms, legs) occupy spatial positions consistent with the reference $\mathbf{x}_{0}$, thereby respecting the original pose. This modified denoising path produces a pose-aligned score estimate $\epsilon^{\text{align}}_\eta(\mathbf{x}_t; y,t,\mathbf{x}_{0,t/2})$. The gradient of the SDS loss computed with this score (Eq.~\ref{eq:sds-with-attention}) therefore carries a signal that penalizes geometric updates, which would break the part-to-part correspondence established by attention. During backpropagation, this gradient $\nabla_\mathcal{J}\mathcal{L}_\text{SDS}^{\text{align}}$ flows through the differentiable renderer $\partial x/\partial \mathcal J$ to update the per-face Jacobians $\mathcal J$. Consequently, the Jacobians are optimized not only for textual alignment but are also regularized by the image-space structural prior enforced by attention sharing, leading to deformations that preserve the source mesh's skeletal pose and part proportions.

While this stage focuses on local detail deformations, the Jacobians may introduce incoherent deformations among adjacent faces. To mitigate this issue, following \cite{gao2023text-deformer}, we introduce a Jacobian regularization loss
that penalizes deviations of the Jacobians from the identity matrix $I$, which is:
\begin{equation}
    \mathcal{L}_\text{reg}=\alpha\sum_{i=1}^{|\mathcal F|}\|J_i-I\|_2,
    \label{eq:jac_reg}
\end{equation} 
where $\alpha$ is a hyper-parameter governing the strength. This regularization effectively bounds the magnitude of per-face deformations, promoting more consistent geometric transformations across neighboring mesh faces.
The total loss $\mathcal{L}_{\text{stage-II}}$ for local detail deformation of this stage is finally defined as:
\begin{equation}
     \mathcal{L}_{\text{stage-II}}=\mathcal{L}_\text{SDS}^{\text{align}}+\mathcal{L}_\text{reg}.
    \label{eq:total_loss}
\end{equation}

\section{Experiments}
\label{sec-experiments}

\subsection{Experimental Setup}
\label{sec:implement}

\textbf{Implementation details.}
The entire two-stage deformation pipeline is executed on a single NVIDIA RTX 6000 Ada GPU with 48 GB of VRAM. During optimization, we process three deformation models simultaneously, each using a batch size of 8 rendered views, resulting in a peak VRAM usage of approximately 44 GB. For rendering, camera parameters are configured as follows: azimuth is uniformly sampled over 360 degrees, elevation follows a beta distribution with a minimum of $0^\circ$, maximum of $60^\circ$, and shape parameters $\alpha = 1.0$ and $\beta = 5.0$. The field of view varies between $30^\circ$ and $90^\circ$, while camera distance ranges from 2.5 to 3.5 units. Lighting is positioned via cosine sampling relative to the camera position, scaled by distance, with an intensity of 5.0. These settings align with TextDeformer to ensure comparability. Under this setting, given a source mesh, our framework can generate three distinct deformation results concurrently within a single optimization run. In both stages, the mesh parameters are optimized using the Adam optimizer, with learning rates set to 2.5e-3 for the Jacobian representation and 2.0e-5 for the Laplacian representation. The first stage converges within 200 iterations and typically requires less than one minute. The second stage jointly optimizes the three deformations for 800 iterations, taking approximately 40 minutes in total. Therefore, the average computation time is about 15 minutes per result.

\textbf{Evaluation metrics.} 
We evaluate our text-driven mesh deformation framework in terms of semantic alignment, pose preservation, and mesh quality. Semantic alignment between the deformed mesh and the input text is measured using the VQA Score \cite{metric:lin2024vqa} and the CLIP Score \cite{radford2021clip}, both computed from $360^\circ$ multi-view renderings with uniformly sampled 180 viewpoints. In particular, we employ CLIP ViT-B/32 for CLIP-based evaluation and CLIP-FlanT5-XXL for VQA-based semantic assessment, enabling both global and fine-grained semantic analysis. Pose preservation is quantified using LPIPS \cite{metric:lpips} (with AlexNet and VGG backbones) between rendered images of the source mesh and those of the deformed mesh, obtained from identical viewpoints; lower scores indicate better pose consistency. We further assess local geometric fidelity by computing the cosine similarity of the Gaussian curvatures between the original and deformed meshes. Finally, mesh quality is evaluated using the self-intersection rate and the average triangle aspect ratio, ensuring that the deformed meshes maintain stable topology and minimal geometric degeneracies.

\textbf{Baselines.} We compare our method with two recent text-guided mesh deformation methods, TextDeformer~\cite{gao2023text-deformer}, and MeshUp~\cite{kim2024meshup}, using public code released by the authors. Specifically, TextDeformer uses CLIP loss to guide Jacobian-based mesh deformation, and MeshUp adopts the Jacobian representation as well but optimizes it with SDS loss.

\subsection{Results and comparisons}
\begin{figure*}[t]
    \centering
    \includegraphics[width=1.0\linewidth]{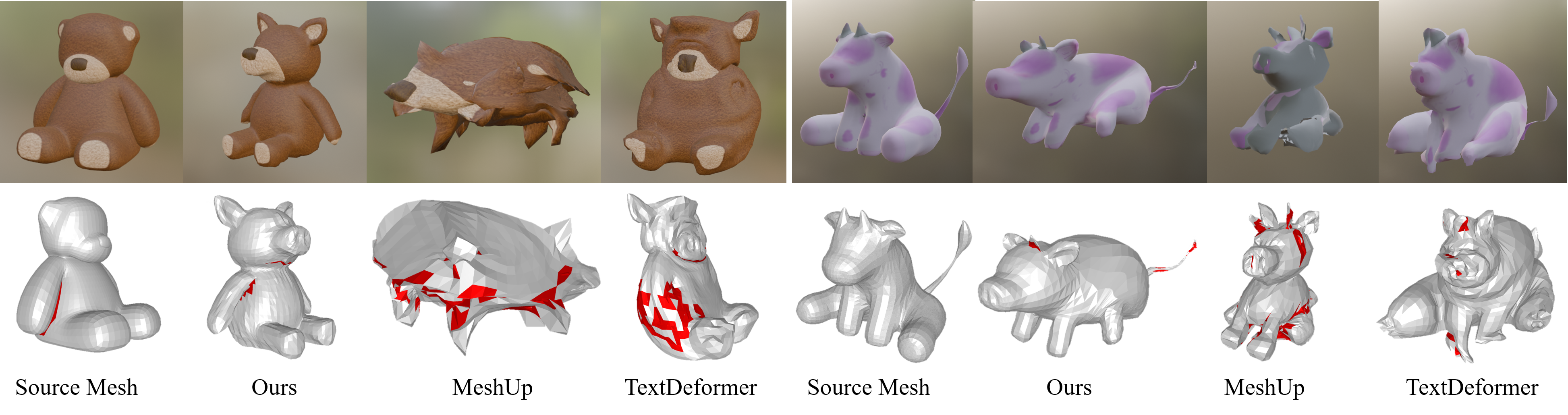}
    \caption{\textbf{Qualitative comparison.} We compare our method against MeshUp and TextDeformer by deforming the same source mesh with the text prompt ``\texttt{a pig}''. The top row visualizes the UV map correspondence and surface distortion by transferring the source mesh's texture onto the deformed results. The bottom row highlights self-intersections in the outputs in red.}
    \label{fig:quality}
\end{figure*}

\textbf{Qualitative Evaluation.} 
Figs. \ref{fig:teaser} and \ref{fig:gallery} present a wide range of deformation results produced by PoseAlign, demonstrating its generalizability and robustness across diverse shape categories. As shown in Fig. \ref{fig:quality}, PoseAlign consistently outperforms existing baselines in terms of pose-faithful deformation and overall mesh quality. In particular, our method better preserves texture coherence and structural correspondence, indicating greater adherence to the source pose during optimization. In contrast, MeshUp relies heavily on diffusion model priors and collapses into a canonical four-legged standing pig pose, completely losing the source model's distinctive sitting posture and introducing noticeable distortions.
TextDeformer attempts to preserve the original pose but exhibits
noticeable stretching artifacts and irregular deformations. These comparisons highlight the effectiveness of our attention-sharing mechanism in maintaining part-to-part correspondence throughout the deformation process.
Furthermore, PoseAlign introduces fewer self-intersections, which are primarily localized around the head and tail regions. In comparison, both MeshUp and TextDeformer exhibit widespread self-intersections in the output meshes, indicating unstable optimization and resulting in geometries unsuitable for downstream applications such as animation or 3D printing.

\begin{table}[ht]
    \centering
    \caption{Quantitative comparison of text-to-mesh alignment.}
    \label{tab:semantic_scores}
    \begin{tabular}{@{}lccc@{}}
        \toprule
         & VQA(↑)          & CLIP(↑)         & \begin{tabular}[c]{@{}c@{}}User Study \\ (Semantic)\end{tabular}  \\ 
        \midrule
        TextDeformer & \textbf{0.7399} & \textbf{0.3129} &  \underline{31.58\%}
                 \\
        MeshUp       & 0.7105          & 0.3035          &  23.68\%
                 \\
        Ours         & \underline{0.7169}          & \underline{0.3050}          &  \textbf{44.74\%}
                 \\
        \bottomrule
    \end{tabular}
\end{table}

\begin{table}[ht]
    \centering
    \caption{Quantitative evaluation of pose preservation.}
    \label{tab:consistency_scores}
    \begin{tabular}{@{}lccc@{}}
        \toprule
             & \begin{tabular}[c]{@{}c@{}}LPIPS- \\ ALEX (↓)\end{tabular} & \begin{tabular}[c]{@{}c@{}}LPIPS- \\ VGG (↓)\end{tabular} & \begin{tabular}[c]{@{}c@{}}User Study \\ (Pose)\end{tabular} \\ 
        \midrule
        TextDeformer & \underline{0.2600}  & \underline{0.1744}   &   9.21\%           \\
        MeshUp       & 0.3545       &  0.2183      &   \underline{13.16\%}          \\
        Ours         & \textbf{0.1542}        & \textbf{0.1036}       &  \textbf{77.63\%}                 \\
        \bottomrule
    \end{tabular}
\end{table}


In summary, PoseAlign not only fulfills the semantic instruction but also realizes a geometrically constrained and stable deformation process. The reduced texture distortion and localized self-intersections indicate that our approach more effectively preserves the input mesh's structural integrity while achieving the target shape transformation, resulting in higher-quality, more reliable 3D assets. 

\textbf{Quantative Evaluation.} 
We also conduct a comprehensive quantitative evaluation comparing our method with TextDeformer and MeshUp along three critical dimensions: semantic alignment, pose preservation, and mesh quality. In addition, we perform a user study focusing on semantic alignment and pose preservation. We presented 30 groups
of mesh results in a random sequence. All evaluations are based on multi-view rendered images to ensure fairness. We have collected 76 valid responses in the user study.

As shown in Tab. \ref{tab:semantic_scores}, TextDeformer achieves the highest scores on both the VQA and CLIP metrics, as expected, since it directly optimizes the mesh using CLIP loss to maximize text-image similarity. In contrast, although our method does not explicitly optimize these semantic metrics, it achieves the highest semantic preference in the user study. This indicates that PoseAlign effectively balances semantic alignment with pose preservation, avoiding over-optimization toward metric scores at the expense of structural fidelity.

Pose preservation and structural consistency are validated in Tab.~\ref{tab:consistency_scores}, where we report LPIPS scores using both AlexNet and VGG backbones. PoseAlign consistently outperforms the two baselines and achieves a substantially higher user-preference rate, demonstrating superior pose fidelity that aligns well with human perception. Finally, Tab.~\ref{tab:mesh_quality_scores} evaluates the intrinsic mesh quality. Our method achieves the second-highest Gaussian curvature similarity while obtaining the best aspect ratio and a low self-intersection rate, producing the overall highest-quality meshes. These results indicate that PoseAlign produces cleaner, more stable, and higher-quality meshes with minimal geometric artifacts, making it more suitable for downstream applications. \revision{Notably, the ``Only Stage I'' row reveals the smoothness vs.\ detail trade-off of our two-stage design: Stage I alone preserves the source curvature most faithfully with zero self-intersections, consistent with its role as a global scaling step that applies Laplacian smoothing without introducing new local features. Stage II then refines the mesh with semantically driven local details via Jacobian optimization, which inevitably alters local curvature but yields the best aspect ratio and a low self-intersection rate, producing the overall highest-quality meshes among all compared methods.}

\begin{table}[htbp]
    \centering
    \caption{Quantitative comparison of mesh quality.}
    \label{tab:mesh_quality_scores}
    \begin{tabular}{@{}lccc@{}}
        \toprule
             & \begin{tabular}[c]{@{}c@{}}Gaussian \\ Curvature(↑)\end{tabular} & \begin{tabular}[c]{@{}c@{}}Aspect \\ Ratio(↓)\end{tabular} & \begin{tabular}[c]{@{}c@{}}Self-Intersection \\ Ratio(↓)\end{tabular} \\ 
        \midrule
        TextDeformer & 0.3218                & \underline{1.8423}          & 6.14\%                      \\
        MeshUp       & 0.2762                & 2.0563          & 10.98\%                    \\
        Ours (Stage I) & \textbf{0.3743}       & 2.0161          & \textbf{0\%}               \\
        Ours (full)        & \underline{0.3562}    & \textbf{1.7082} & \underline{3.82\%}         \\
        \bottomrule
        \end{tabular}
\end{table}


\subsection{Ablation study}

\begin{figure*}[ht]
    \centering
    \includegraphics[width=0.8\linewidth]{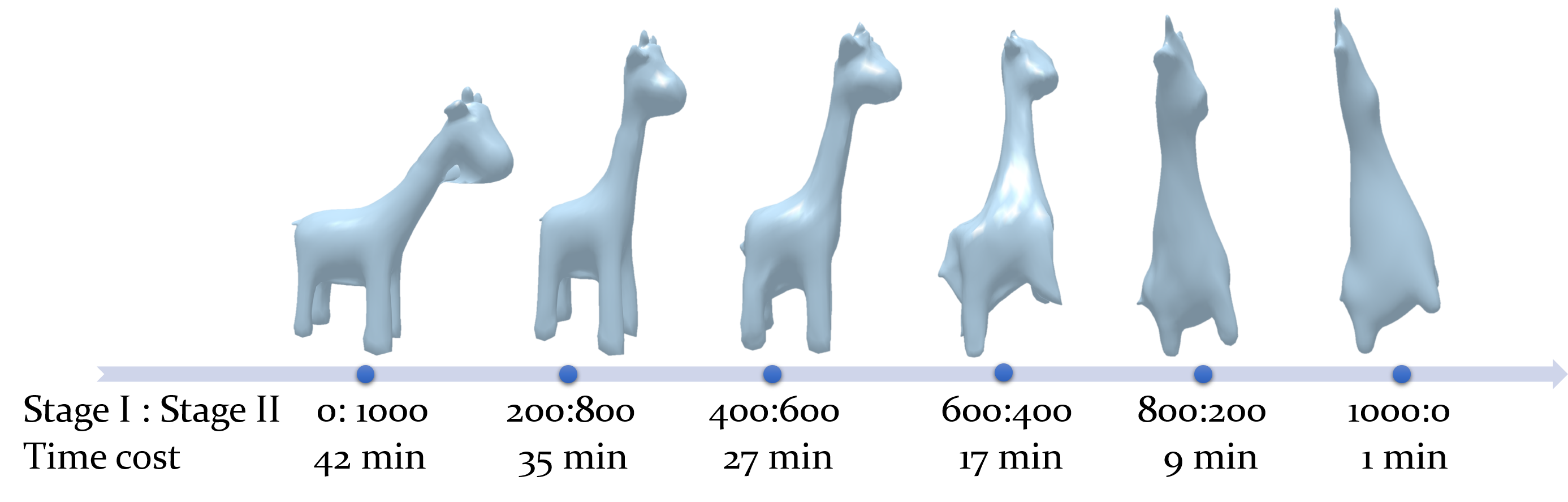}
    \caption{Visual comparison under different Stage I and Stage II optimization iteration ratios.}
    \label{fig:stage1-vs-stage2}
\end{figure*}

\begin{figure}[ht]
    \centering
    \includegraphics[width=1.0\linewidth]{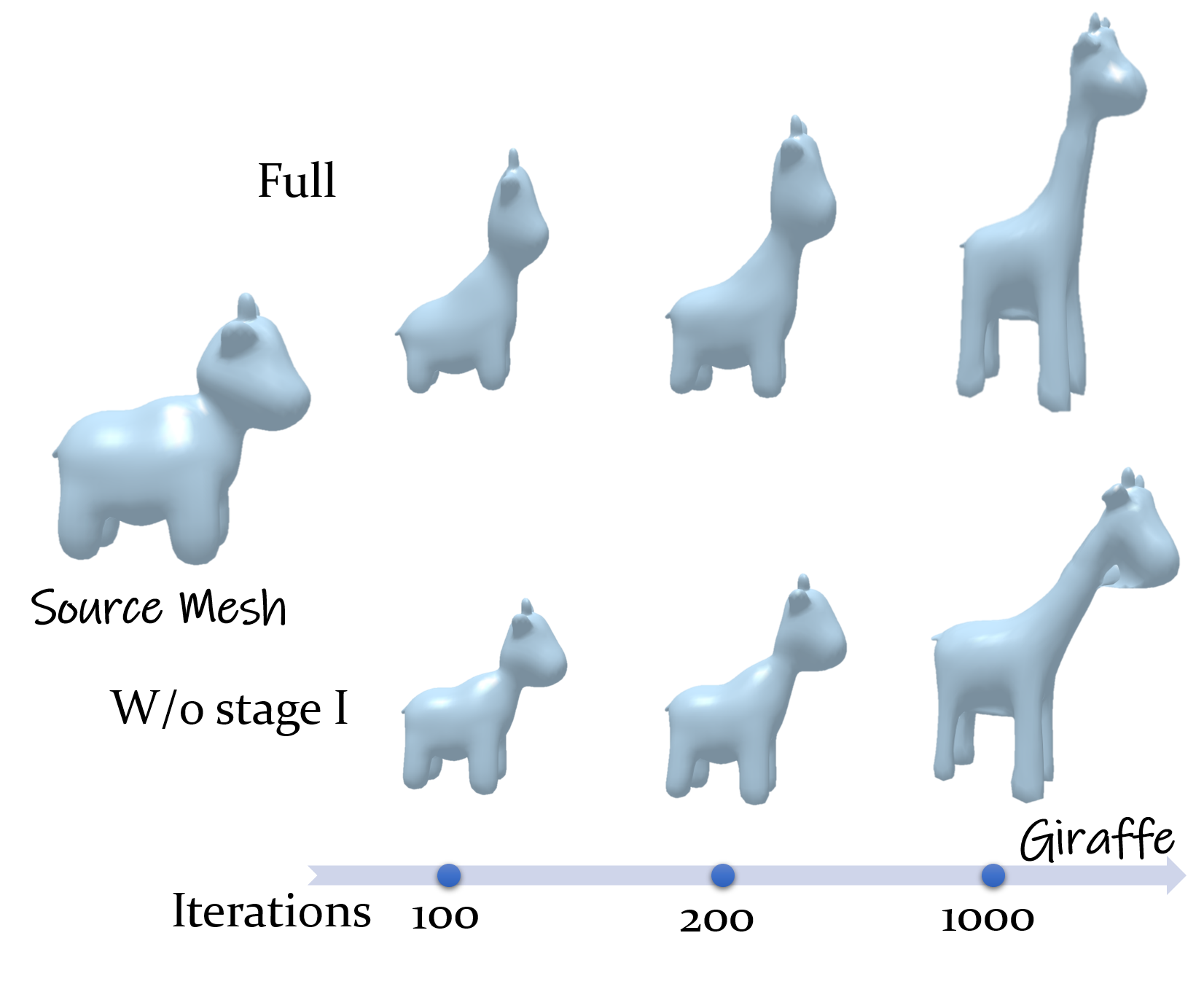}
    \caption{\textbf{Necessity of Stage I.} The top row illustrates the deformation process of our full method, where the source mesh undergoes a smooth and controlled transformation into a target giraffe shape across 100, 200, and 1000 iterations. The bottom row shows the results when Stage I is omitted.} 
    \label{fig:abl:stage1}
\end{figure}

\textbf{Stage I.} To verify the indispensable role of the first stage in our proposed two-stage framework, we conduct a comprehensive ablation study on Stage I. 
As shown in Fig. \ref{fig:stage1-vs-stage2}, we first evaluate different combinations of optimization iterations for Stage I and Stage II. Empirically, the setting with 200 iterations for Stage I and 800 iterations for Stage II yields the best visual quality while achieving faster deformation than the pipeline without Stage I. Then, as illustrated in Fig.~\ref{fig:abl:stage1}, the full method exhibits coherent and progressive deformation, with clear neck elongation and leg refinement emerging by 100 iterations. In contrast, the ablated variant shows minimal deformation at the same stage and only partial shape development by 200 iterations, indicating slower convergence. After 1000 iterations, the full pipeline produces a geometrically consistent giraffe with smooth surfaces and well-proportioned limbs, whereas the ablated result exhibits an irregular neck geometry near the chin. These artifacts highlight the importance of Stage I in providing a robust geometric initialization, which is crucial for high-quality and stable refinement in Stage II.

\begin{figure}[!t]
    \centering
    \includegraphics[width=1.0\linewidth]{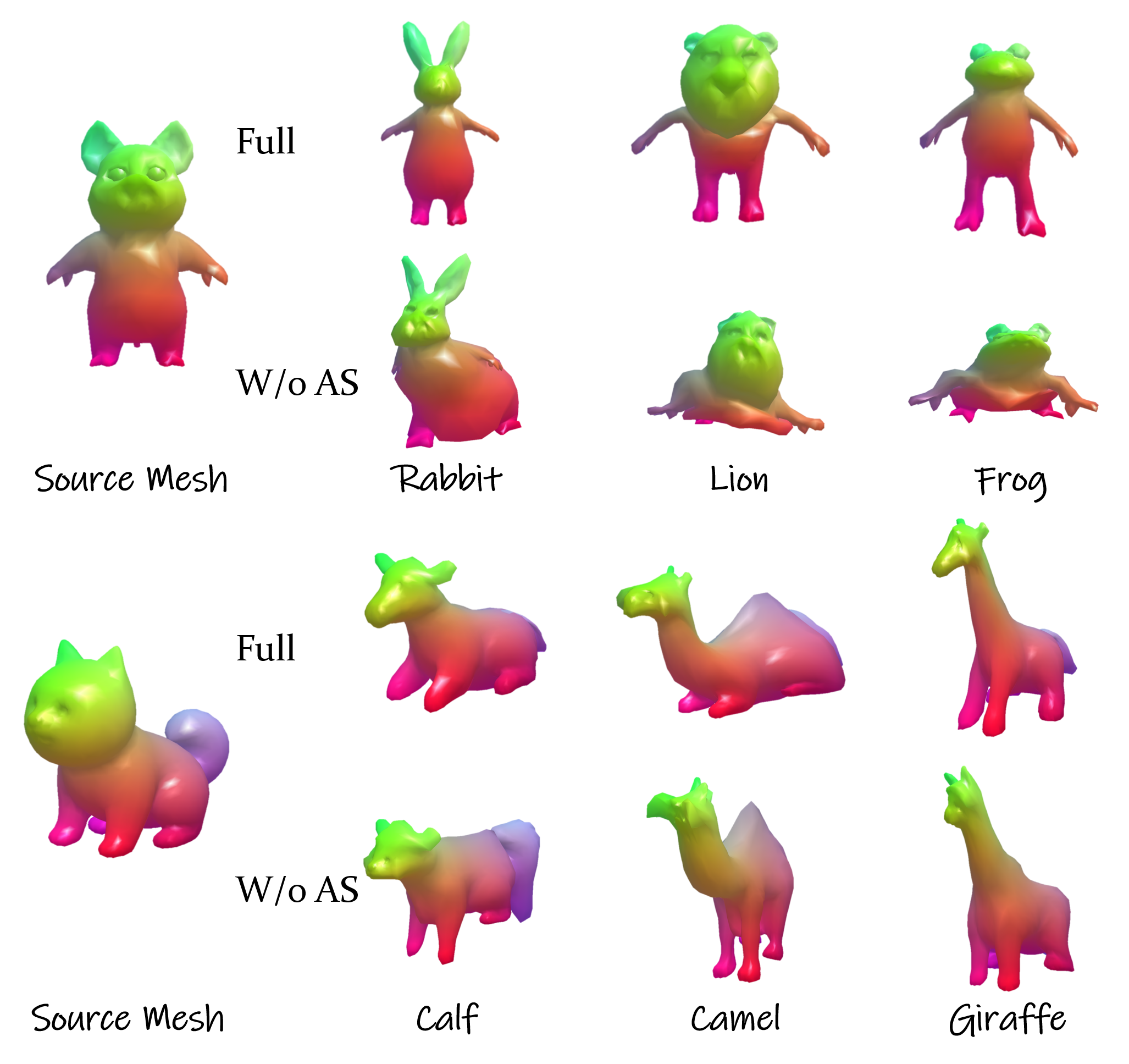}
    \caption{\textbf{Ablation on attention sharing.} All deformations are guided by identical prompts, with the same settings: 200 iterations for Stage I and 800 for Stage II.}
    \label{fig:abl:sa}
\end{figure}

\textbf{Attention Sharing.} We also conduct an ablation study to evaluate the effect of the attention-sharing (AS) mechanism in our pose-aligned SDS loss. As presented in Fig. \ref{fig:abl:sa}, 
under identical text prompts and training budgets, the full method successfully deforms both displayed meshes to satisfy the semantic targets while robustly preserving their original poses, regardless of whether the pose is bipedal or squatting. In contrast, removing the AS results in significant pose degradation, leading to noticeable limb disorganization and torso distortion. These results clearly demonstrate that attention sharing is essential for maintaining pose consistency and part-to-part correspondence during text-driven mesh deformation.

\subsection{Additional results and applications}

\begin{figure}[!t]
    \centering
    \includegraphics[width=1.0\linewidth]{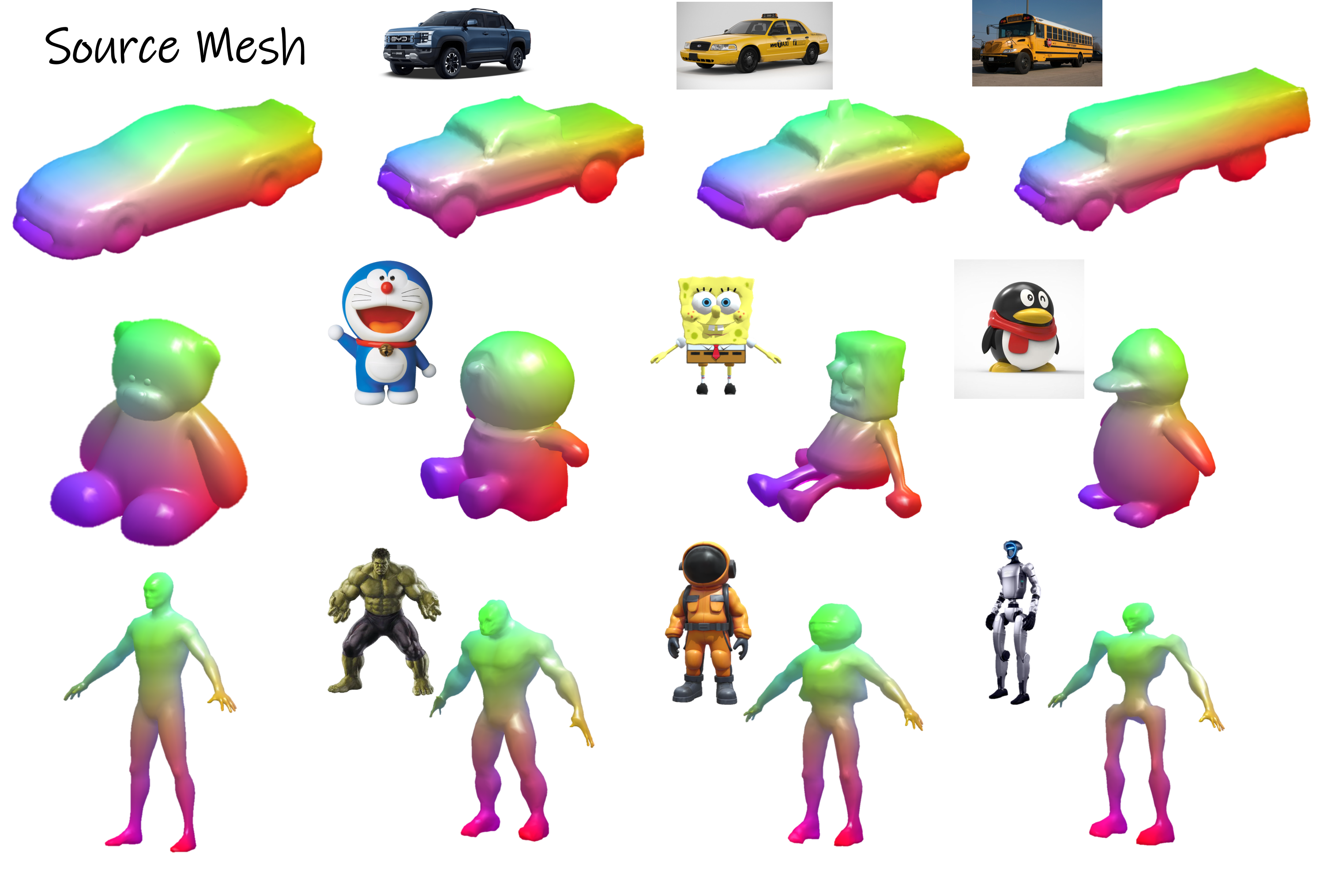}
    \caption{We use the textual inversion technique to condition the text-to-image model with images represented by their inverted textual tokens.}
    \label{fig:inversion}
\end{figure}
    
\textbf{Image-Guided Deformation.} We first extend our framework to support image-guided mesh deformation by incorporating the Textual Inversion \cite{relate:person:textualinversion} technique. 
A reference image is encoded into the text embedding space via textual inversion, enabling the source mesh to be deformed under the guidance of an image-defined concept.
This is particularly beneficial for concepts with substantial shape variations, such as different vehicle types, or for targets that are difficult to describe precisely in text, e.g., cartoon characters. Experimental results shown in Fig. \ref{fig:inversion} demonstrate that our method effectively translates image-based visual cues into coherent geometric deformations while preserving the core structural attributes of the source mesh. This extension significantly enhances the flexibility and applicability of our approach in practical scenarios where visual references are more intuitive than textual descriptions.

\begin{figure}[!t]
    \centering
    \includegraphics[width=0.8\linewidth]{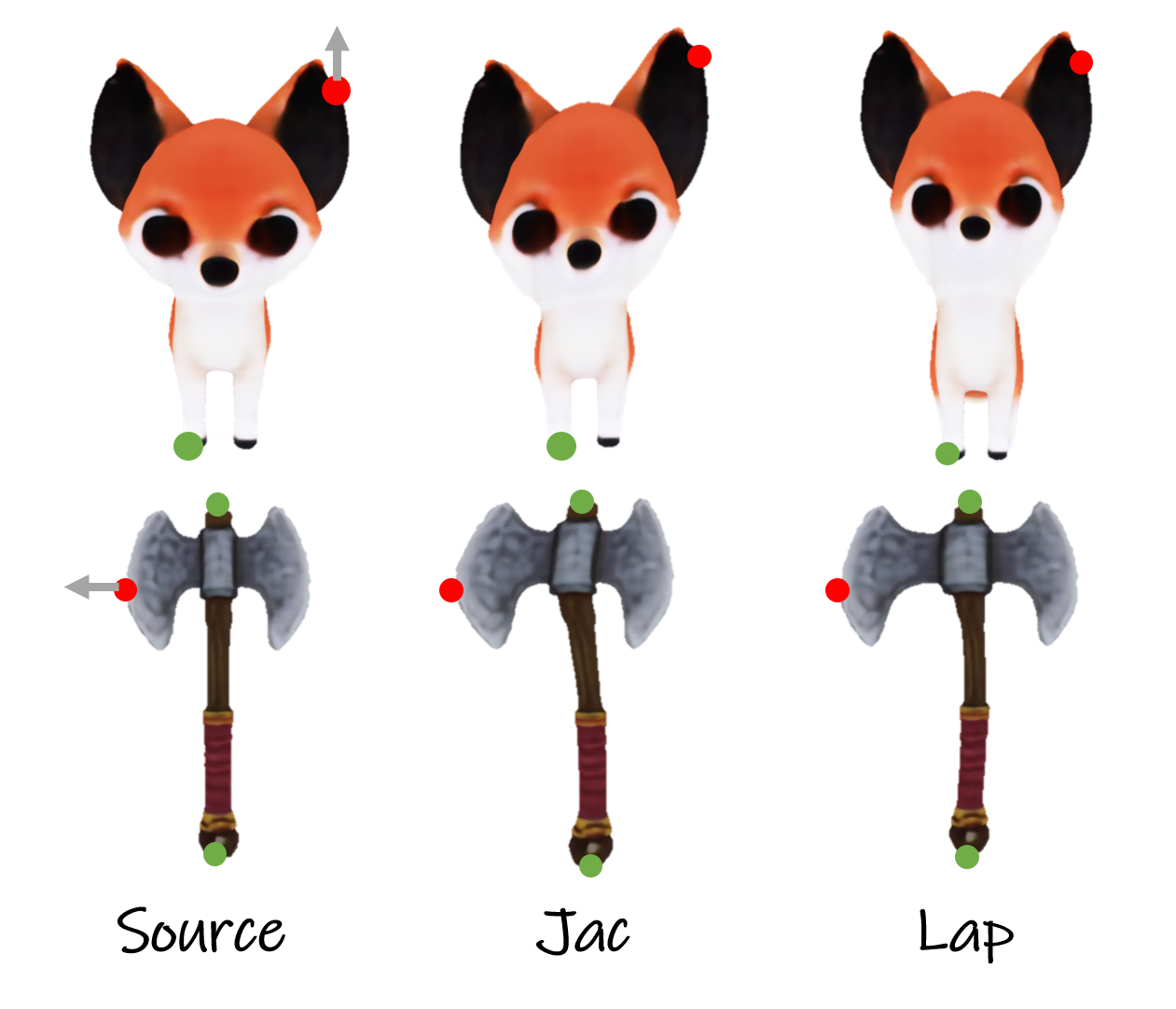}
    \caption{We visualize the results of drag-based mesh deformation after 300 iterations of optimization, using the Jacobians (``Jac'') and the Laplacians (``Lap'') as geometric representations, respectively. The deformation is guided by the red handle points, with the edit direction indicated by a gray arrow, and green points as anchors.}
    \label{fig:handle}
\end{figure}

\textbf{Drag-Based Deformation.} We then apply the Laplacians to the drag-based deformation task and also compare the results with those using the Jacobians, as illustrated in Fig. \ref{fig:handle}. In the top row, a cartoon-style fox model with its right ear pulled upward, the Jacobian-based deformation produces disproportionate ear distortions, whereas the Laplacian-based approach preserves natural symmetry, with both ears rising uniformly and the face remaining horizontally consistent. Similarly, in the bottom row, for an axe model whose left blade is pulled to the left, the Jacobian method introduces unnatural stretching around the handle, whereas the Laplacian method produces a smoother, more coherent deformation that distributes changes evenly, including the slight stretching of the opposite blade. The experimental findings show that Laplacians excel at preserving global structural integrity and minimizing artifacts, making them well-suited for applications requiring smooth, symmetric shape manipulation.

\begin{figure*}[!t]
    \centering
    \includegraphics[width=1.0\linewidth]{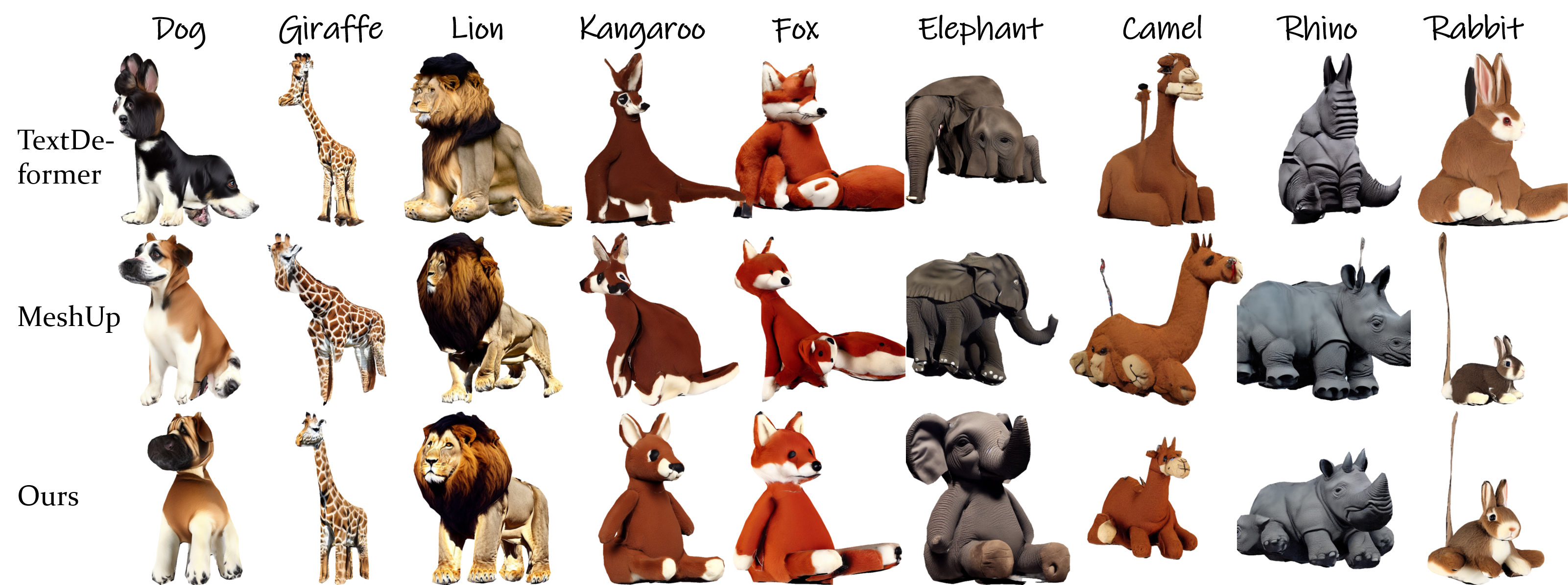}
    \caption{\textbf{Deformed meshes with generated textures.} The results are organized into three groups according to the source mesh used. Specifically, the first group (column 1-3) uses the one shown in the first row of Fig. \ref{fig:gallery}, while the middle (column 4-6) and right (column 7-9) groups use the source meshes shown in the second and third rows of Fig. \ref{fig:gallery}, respectively.}
    \label{fig:texture}
\end{figure*}

\textbf{Texture Application.} We further evaluate the practicality of the deformed meshes in downstream applications by applying SyncMVD~\cite{app:Syncmvd} for texture generation. SyncMVD employs a synchronized multi-view diffusion approach to create consistent and seamless textures from text prompts, making it well-suited for assessing mesh quality after deformation. As shown in Fig. \ref{fig:texture}, thanks to the better semantic alignment, meshes produced by our method exhibit more natural, pose-consistent geometries, such as the proportionally sized limbs in animal models. This enables SyncMVD to generate visually cohesive textures, with enhanced detail alignment and fewer artifacts. In contrast, baseline methods often introduce distortions that lead to texture misalignment or inconsistencies, as seen in irregular stretching and seams. These results demonstrate that PoseAlign yields higher-quality geometries that integrate more effectively with advanced texture-synthesis pipelines, underscoring its effectiveness as a preprocessing step for downstream 3D content-creation tasks.

\subsection{Limitations}

\begin{figure*}[!t]
    \centering
    \includegraphics[width=0.8\linewidth]{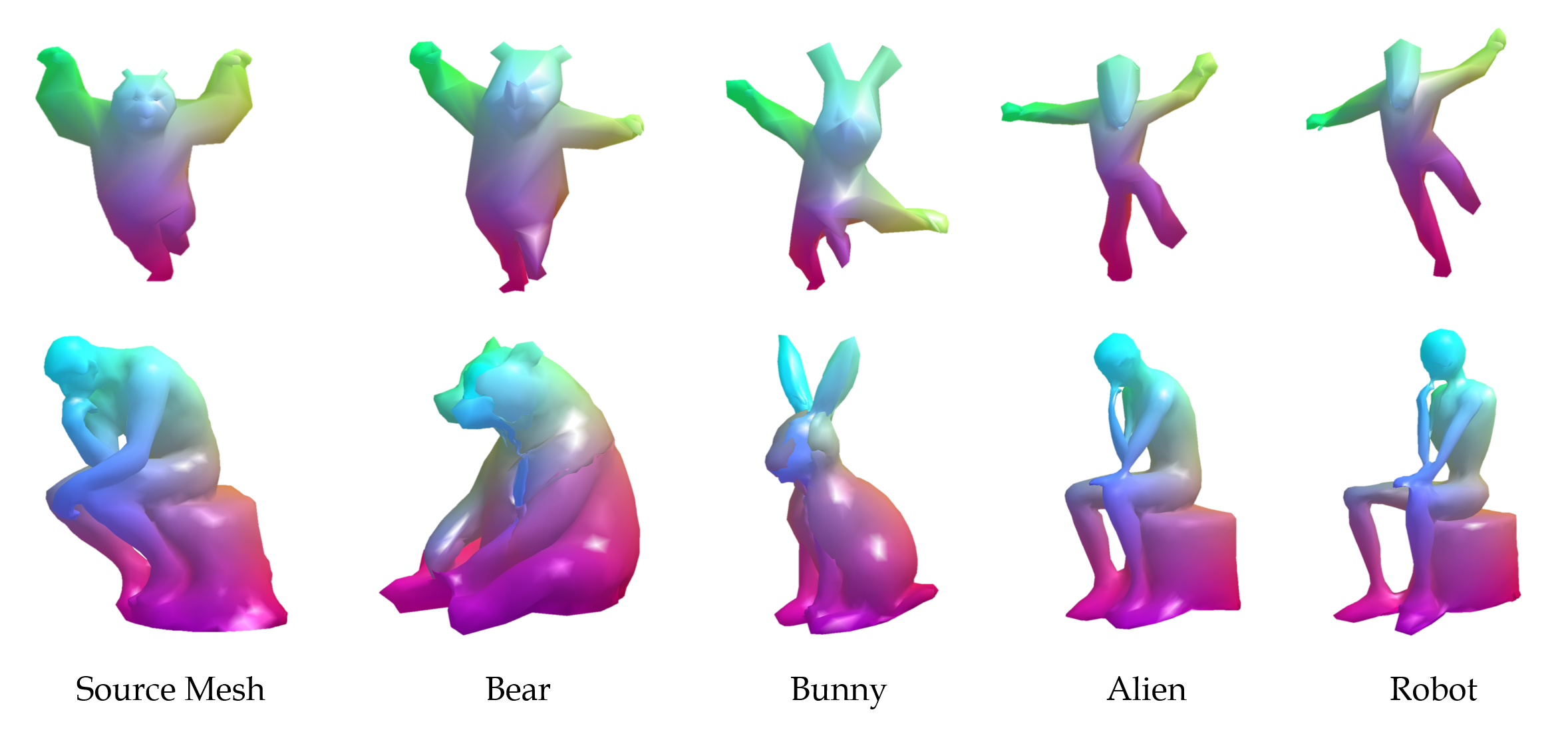}
    \caption{\textbf{Validation of PoseAlign under challenging poses.} Mesh Deformation of Kung Fu Panda and The Thinker. The results highlight our method's limitations in preserving intricate source poses during deformation, particularly for human-to-animal transformations.}
    \label{fig:complex-pose}
\end{figure*}

Despite the promising results achieved by PoseAlign, several limitations remain and point to directions for future work. First, the computational efficiency of the framework is constrained by the optimization in Stage II. On average, Stage II requires approximately 15 minutes per deformation, primarily due to the iterative optimization of the pose-aligned SDS loss with diffusion models. This limits PoseAlign's applicability in time-critical or interactive scenarios. Future work may investigate more efficient optimization schemes or lightweight diffusion models to reduce the computational overhead. 

Second, the performance of our method is sensitive to the quality of the input mesh. Noisy geometry or poor topology in the source mesh can be amplified during deformation, potentially degrading the final results and limiting generalization. Incorporating mesh preprocessing or topology-aware regularization could help mitigate this issue. 

Moreover, the efficacy of PoseAlign is constrained by the semantic priors of the underlying 2D diffusion model, especially when handling highly complex source poses. As illustrated in Fig.~\ref{fig:complex-pose}, our method encounters two primary challenges in such scenarios: First, when deforming a complex human pose into an animal, the distinctive source pose is often lost. 
\revision{We attribute this primarily to the training bias of the underlying diffusion model---animals are overwhelmingly depicted in canonical poses in the training data, so the model lacks priors for non-canonical configurations (e.g., Kung Fu Panda). Futhermore, when the semantic gap between source and target is large (e.g., the Thinker to animal), the diffusion model fails to establish correct semantic correspondences between the two domains, which undermines the attention-sharing mechanism's ability to propagate pose information. Since our attention-sharing mechanism propagates whatever the diffusion model produces, this bias directly affects the geometric supervision. }
Second, for human-to-human deformation, our method struggles to control geometry at a fine-grained level and may not produce results that visually distinguish the output from the source mesh. \revision{This is a genuine limitation of our method: when the source and target share a similar skeletal structure, the SDS-induced geometric changes may be too subtle to yield perceptually distinct results.} Addressing these limitations by integrating stronger 3D-aware or pose-conditioned diffusion priors is a promising direction for future work.
\section{Conclusion}

In this paper, we present PoseAlign, a novel framework for text-guided mesh deformation that effectively preserves the original mesh pose while aligning with semantic prompts. By decomposing the deformation process into a Global Pose Scaling stage followed by a Local Detail Sculpting stage, our method achieves a superior balance between deformation controllability, semantic alignment, geometric fidelity, and mesh quality. Extensive experiments demonstrate that our approach consistently outperforms state-of-the-art methods, including TextDeformer and MeshUp, across multiple evaluation metrics. These results highlight the effectiveness and robustness of PoseAlign in producing high-quality, pose-faithful 3D deformations.

\begin{acknowledgements}
This work was supported in part by ICFCRT (W2441020), Guangdong Basic and Applied Basic Research Foundation (2023B1515120026), SZU Teaching Reform Research Project (JG2026011), and Scientific Development Fund from Guangdong Provincial Key Laboratory of Visual Media and Multidimensional Intelligence.
\end{acknowledgements}

\bibliographystyle{spmpsci}      
\bibliography{mybibfile.bib}   



\newpage
\appendix

\newpage

\appendix

\newcommand{\secondtitle}[1]{
    {
        \centering
        \normalfont
        \Large\textbf{#1}
        \vspace{1\baselineskip}
    }
}

\secondtitle{
    {
        Appendix
    }
}

\section{Pseudo-code}

\begin{figure}[h]
    \centering
    \resizebox{1.0\linewidth}{!}{%
        \begin{minipage}{\linewidth}
        
        \begin{algorithm}[H]
            \caption{Pseudo-code of PoseAlign}
            \label{alg-pseudo-code}
            \begin{algorithmic}[1]
                \Require Source mesh $\mathcal{M}_0 = \{\mathcal{V}_0, \mathcal{F}_0\}$, text prompt $y$
                \Ensure Deformed mesh $\mathcal{M}$
                
                \State \textbf{Stage I: Global Pose Scaling}
                \State Initialize Laplacian coordinates $\mathcal{L}$ for $\mathcal{M}_0$ via Eq. (4)
                \For{iteration $= 1$ to $N_1$ (e.g., 200)}
                    \State Solve Poisson equation for deformation map $\phi^*$ via Eq. (5)
                    \State Compute deformed mesh $\mathcal{M}$ from $\phi^*$
                    \State Render multi-view images $x$ from $\mathcal{M}$ using differentiable rendering
                    \State Compute CLIP loss
                    \State Update $\mathcal{L}$ via gradient descent: $\nabla_{\mathcal{L}} \mathcal{L}_{\text{CLIP}}$
                \EndFor
                \State Output of Stage I $\mathcal{M}_{\text{stageI}}$ 
                
                \State \textbf{Stage II: Local Detail Deformation}
                \State Initialize Jacobians $\mathcal{J}$ for $\mathcal{M}_{\text{stageI}}$ via per-face gradients
                \For{iteration $= 1$ to $N_2$ (e.g., 800)}
                    \State Solve Poisson equation for vertex positions via Eq. (1)
                    \State Compute updated mesh $\mathcal{M}$ from $\phi^*$
                    \State Render images $x$ from $\mathcal{M}$ and $x_0$ from $\mathcal{M}_0$ using differentiable rendering
                    \State Perturb images: $x_t = \text{noise}(x, t)$, $x_{0,t/2} = \text{noise}(x_0, t/2)$
                    \State Compute Align SDS loss via Eq. (8)
                    \State Compute Jacobian regularization loss via Eq. (9)
                    \State Compute total loss
                    \State Update $\mathcal{J}$ via gradient descent: $\nabla_{\mathcal{J}} \mathcal{L}_{\text{stageII}}$
                \EndFor
                \State \Return Final deformed mesh $\mathcal{M}$  
            \end{algorithmic}
        \end{algorithm}
        
        \end{minipage}
    }
\end{figure}

\section{Proof of Eq.\ref{eq:lap_poisson}}
\begin{thm}
    If $u$ is a function that minimizes $$E(u)=\int_{\Omega}\|\Delta u-L\|^2 $$ over a given domain $\Omega$, then $u$ satisfies $\Delta^2 u=\Delta L$.
    \label{state2}
\end{thm}
\begin{proof}[Proof]
    We consider a perturbation $u_\epsilon=u+\epsilon w$, where $w$ satisfies $w\in \mathbb{C}^2, w|_{\partial\Omega}=0, \nabla w|_{\partial\Omega}=0$.
    We can get the energy of the perturbed function:
    \begin{equation}
        E(u_\epsilon)=\int_\Omega\left(\Delta(u+\epsilon w)-L\right)^2\mathrm{d}x=\int_\Omega(\Delta u+\epsilon\Delta w-L)^2\mathrm{d}x.
        \label{eq:enery-of-perturbed-function}
    \end{equation}
    Differentiate with respect to $\epsilon$ and evaluate at $\epsilon=0$, we get
    $$\delta E=\frac{\mathrm{d}}{\mathrm{d}\epsilon}E(u_\epsilon)\bigg|_{\epsilon=0}=\int_\Omega2(\Delta u-L)\Delta w \mathrm{d}x.$$
    Since $u$ is a minimizer, $\delta E=0$ for all admissible $w$. Thus,
    $$\int_\Omega (\Delta u-L)\Delta w \mathrm{d}x=0.$$
    Using integration by parts, we have
    $$\int_{\Omega}\Delta u\cdot\Delta w\mathrm{d}x=-\int_{\Omega}\nabla\Delta u\cdot\nabla w\mathrm{d}x+\text{boundary terms},$$
    $$\int_{\Omega}L\Delta w\mathrm{d}x=-\int_{\Omega}\nabla L\cdot\nabla w\mathrm{d}x+\text{boundary terms}.$$
    As $\nabla w$ vanish on the boundary, the boundary terms equal to zero. Using integration by parts again:
    $$\int_{\Omega}\nabla\Delta u\cdot\nabla w\mathrm{d}x=-\int_{\Omega}\Delta^2 u \cdot w\mathrm{d}x+\text{boundary terms},$$
    $$\int_{\Omega}\nabla L\cdot\nabla w\mathrm{d}x=-\int_{\Omega}\Delta L\cdot w\mathrm{d}x+\text{boundary terms}.$$
    Again, as $w$ vanish on the boundary, we have
    $$\int_{\Omega}w\cdot(\Delta^2 u-\Delta L)\mathrm{d}x=0.$$
    Since $w$ is arbitrary, we conclude 
    $$\Delta^2u=\Delta L.$$
\end{proof}

Statement \ref{state2} reveals that we can get the deformation function through a Laplace matrix, \ie $u=\Delta^{-1} L$.

\end{document}